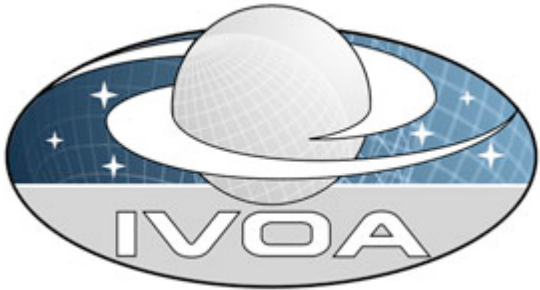

**I**nternational

**V**irtual

**O**bservatory

**A**lliance

# IVOA Registry Interfaces
# Version 1.0

## IVOA Recommendation 2009 November 4

**Authors:**
  Kevin Benson, Ray Plante, Elizabeth Auden, Matthew Graham, Gretchen Greene, Martin Hill, Tony Linde, Dave Morris, Wil O'Mullane, Guy Rixon, Aurélien Stébé, Kona Andrews

---

## Abstract

Registries provide a mechanism with which VO applications can discover and select resources—e.g. data and services—that are relevant for a particular scientific problem.  This specification defines the interfaces that support interactions between applications and registries as well as between the registries themselves.  It is based on a general, distributed model composed of so-called *searchable* and *publishing* registries.  The specification has two main components:  an interface for searching and an interface for *harvesting*.  All interfaces are defined by a standard Web Service Description Language (WSDL) document; however, harvesting is also supported through the existing Open Archives Initiative Protocol for Metadata Harvesting, defined as an HTTP REST interface.  Finally, this specification details the metadata used to describe registries themselves as resources using an extension of the VOResource metadata schema.

## Status of This Document

This is an IVOA Recommendation. The <u>first release of this document</u> was 2006 October 15. Note that a previous version of this document featured the version number 1.01 in compliance with IVOA Document Standards v1.0 [STDv1.0]. This version, v1.0, adopts the recommendation for version numbering of IVOA Document Standards v1.2 [STDv1.2].

This document has been produced by the IVOA Resource Registry Working Group. It has been reviewed by IVOA members and other interested parties and has been endorsed by the IVOA Executive Committee as an IVOA Recommendation. It is a stable document and may be used as reference material or cited as a normative reference from another document. IVOA's role in making the Recommendation is to draw attention to the specification and to promote its widespread deployment. This enhances the functionality and interoperability inside the Astronomical Community.

A list of current IVOA Recommendations and other technical documents can be found at http://www.ivoa.net/Documents/.

## Acknowledgements

This document has been developed with support from the National Science Foundation's Information Technology Research Program under Cooperative Agreement AST0122449 with The Johns Hopkins University, from the UK Particle Physics and Astronomy Research Council (PPARC), and from the European Commission's Sixth Framework Program via the Optical Infrared Coordination Network (OPTICON).

## Conformance-related definitions

The words "MUST ", "SHOULD", "MAY", "RECOMMENDED", and "OPTIONAL" (in upper or lower case) used in this document are to be interpreted as described in IETF standard, RFC 2119 [RFC 2119].

The **Virtual Observatory (VO)** is a general term for a collection of federated resources that can be used to conduct astronomical research, education, and outreach. The **International Virtual Observatory Alliance (IVOA)** is a global collaboration of separately funded projects to develop standards and infrastructure that enable VO applications.

A **Web Service** (when capitalized as it is here) refers to a service that is in actuality described by a Web Service Description Language (WSDL) [WSDLv1.1] document.

> **Note:**
> Boxed comments labeled "Note," such as this one, are intended to provide tips and further explanation of how the standard is expected to be used or implemented. Their contents are informative, rather than normative—that is, they are not technically part of the standard

# Contents

# 1   Introduction

In the Virtual Observatory (VO), registries provide a means for discovering useful data and services.  To make discovery efficient, a registry typically represents to some extent a centralized warehouse of resource descriptions; however, the source of this information—the resources themselves and the data providers that maintain them—are distributed.  Furthermore, there need not be a single registry that serves the entire international VO community.  Given the inherent distributed nature of the information used for resource discovery, there is clearly a need for common mechanisms for registry communication and interaction.

This document describes the standard interfaces that enable interoperable registries.  These interfaces are based in large part on a Web Service definition in the form of a WSDL document [WSDLv1.1], which is included in this specification.  Through these interfaces, registry builders have a common way of sharing resource descriptions with users, applications, and other registries.  Client applications can be built according to this specification and be able to discover and retrieve descriptions from any compliant registry.

This specification does not preclude a registry builder from providing additional value-added interfaces and capabilities.  In particular, they are free to build interactive, end-user interfaces in any way that best serves their target community.  In a similar spirit, this specification does not intend to enforce completely identical behavior of required operations across all compliant implementations.  In particular, this specification does not require that identical search queries sent to different compliant registries return identical results.  Implementations are free to support different strategies for evaluating an ambiguous query (such as a keyword search) and ordering the results in a way that best serves the target community.

## 1.1   Registry Architecture and Definitions

A **registry** is first a repository of structured descriptions of *resources*. In the VO, a **resource** is defined by the IVOA Recommendation, "Resource Metadata for the Virtual Observatory" (RM) [Hanisch 2004]:

> A *resource* is a general term referring to a VO element that can be described in terms of who curates or maintains it and which can be given a name and a unique identifier.  Just about anything can be a resource: it can be an abstract idea, such as sky coverage or an instrumental setup, or it can be fairly concrete, like an organization or a data collection.

Organizations, data collections, and services can be considered as classes of resources. The most important type of resource to applications is a service that actually does something. What is available at a particular resource is described through the content of metadata, whereas the service metadata describes how to access it. The RM describes a registry, then, as "a service for which the response is a structured description of resources" [Hanisch 2004]. Each resource description it returns is referred to as a **resource record**.

This specification is based on the general IVOA model for registries [Plante 2003], which builds on the RM model for resources. In the registry model, the VO environment features different types of registries that serve different functions. The primary distinction is between *publishing* registries and *searchable* ones. A secondary distinction is *full* versus *local*.

A **searchable registry** is one that allows users and client applications to search for resource records using selection criteria against the metadata contained in the records. The purpose of this type of registry is to aggregate descriptions of many resources distributed across the network. By providing a single place to locate data and services, applications are saved from having to visit many different sites to just to determine which ones are relevant to the scientific problem at hand. A searchable registry gathers its descriptions from across the network through a process called *harvesting*.

A **publishing registry** is one that simply exposes its resource descriptions to the VO environment in a way that allows those descriptions to be harvested. The contents of these registries tend to be limited to resources maintained by one or a few providers and thus are local in nature; for example, a data center will run its own publishing registry to expose all the resources it maintains to the VO environment. Since the purpose is simply publishing and not to serve users and applications directly, it is not necessary to support full searching capabilities. This simplifies the requirements for a publishing registry: not only does it not need to support the general search interface, the storage and management of the records can be simpler. While a searchable registry in practice will necessitate the use of a database system, a publishing registry can easily store its records as flat files on disk.

Note that some registries can play both roles; that is, a searchable registry may also publish its own resource descriptions.

A secondary distinction is *full* versus *local*. A **full registry** is one that attempts to contain records of all resources known to the VO. In practice, this attribute is associated only with searchable registries, as in the so-called **full searchable registry**. It is expected that there will be several such registries available, perhaps each run by a major VO project; this not only avoids the single point of failure, but also allows some specialization to serve the particular needs of the project that maintains it. A **local registry**, on the other hand, contains only a subset of known resources.  In practice, all publishing registries are local; however, we expect that there may be **local searchable registries** that specialize in particular types of resources, perhaps oriented toward a scientific topic.

As mentioned above, **harvesting** is the mechanism by which a registry can collect resource records from other registries.  This mechanism is used by full searchable registries to aggregate resource records from many publishing registries.  It can also be used to synchronize two registries to ensure that they have the same contents.  Harvesting, in this specification, is modeled as a "pull" operation between two registries.  The **harvester** refers to the registry that wishes to receive records (usually a searchable registry); it sends its request to the **harvestee** (usually the publishing registry), which responds with the records. Harvesting is intended to be a much simpler process than search and retrieval; nevertheless, there are at least two kinds of filtering that a harvestee needs to support:

- **Filtering by date:**  this allows the harvester to return to the harvestee periodically to retrieve only new and updated records.
- **Filtering by ownership:**  harvesting only those records that originated with the harvestee (as opposed to those that may have been harvested from other registries) prevents a harvester from receiving duplicate records from multiple registries.

Other kinds of filtering can be useful as well (such as filtering on resource type). Note, however, that filtering is not intended to be an equivalent to arbitrary searching; rather, it is a gross selection that can be easily implemented without having to process the contents of each record.

## 1.2  Specification Summary

The purpose of the *registry* is to be used by other applications to provide access to various types of resources. At the programmatic level, connectivity between the registry and other applications is ensured through the registry interface as defined by this document. Much of the interface is defined as a SOAP-based Web Service and is described the WSDL documents included in Appendix A.1; however, the harvesting interface (section 3) is specifically defined by the Open Archives Initiative standard called the Protocol for Metadata Harvesting (OAI-PMH) [OAI].   To define the registry interface, this document includes the following definitions:

- —  The meaning and behavior of four required searching operations, one optional search operation, and six required harvesting operations.
- —  The required input arguments for each operation.
- —  The XML Schema used to encode response messages.
- —  The meaning of the output for each operation.

The registry interface is composed of two independent parts. The harvesting interface provides the mechanism for registries to talk to each other and share information. The searching interface is used by clients that want to discover resources to use as part of a VO application. A registry may implement either interface or both, depending on the roles it intends to play.

The searching interface can return XML descriptions of resources, or resource records, and it consists of four required operations, described in more detail in section 2:

- **Search** returns active resource records that match a specific set of constraints.
- **KeywordSearch** returns active resource records containing specified keywords.
- **GetResource** returns a single resource records identified by its unique IVOA identifier.
- **GetIdentity** returns the resource record describing the searchable registry itself.

The searching interface also includes an optional, alternative search option based on XQuery.

The harvesting interface, which allows one registry to collect resource records from other registries, leverages the OAI-PMH standard [OAI]. Described in more detail in section 3, the OAI-PMH interface is composed of six operations. The most commonly used harvesting operation is the **ListRecords**, which can return the descriptions of the resources of a particular category, or *set*, that have been created or updated since a specified date. Normally the harvester will request records from the set that originate from that registry (as opposed to those harvested from another registry). This set is said to be *managed* by the registry. The complete list of OAI-PMH harvesting operations is:

- **Identify** returns the resource record describing the harvestable registry itself.
- **ListIdentifiers** lists the identifiers of records that have changed since a given date
- **ListRecords** lists the full descriptions of resources that have changed since a given date.
- **GetRecord** returns a single record identified by its identifier.
- **ListMetadataFormats** lists the available output description formats.
- **ListSets** lists the categories of records that can be requested.

The searching and harvesting operations that return resource descriptions do so using the VOResource XML Schema and any of its legal extensions [VOResource].

An IVOA-compliant registry can itself be described in a registry using the Registry VOResource extension described in section 4. To be considered an IVOA-compliant searchable registry, the registry must support the search interface (section 2). An IVOA-compliant publishing registry must support the harvesting interface (section 3). To be considered an IVOA-compliant registry in general, the registry must support the search interface *or* the harvesting interface *or both*.

# 2  Searching

The four required operations that make up the searching interface fall into two groups: *search* and *resolve*. The *search* operations—**Search** and **KeywordSearch**—return a list of one or more resource descriptions held by the registry that matches the input selection criteria. The **Search** operation supports constraint-based searching for resources by means of a query using the Astronomical Data Query Language (ADQL) [Appendix A.5], **KeywordSearch** provides keyword-based searching for resources whose descriptions contain words in an input string.

The resolve operations---**GetResource** and **GetIdentity**---each return one and only one resource description. The **GetResource** resolves a unique IVOA Identifier [Identifier] to the description of the associated resource. The **GetIdentity** returns the resource record of the searchable registry itself.

> **Note:**
> It is important to note that search operations do not support resource harvesting described in section 3. Normally, an end-user would use search to retrieve resource descriptions, but not to selectively harvest information between registries.

These four operations are defined by the WSDL document given in Appendix A.1. Searchable registries must implement all four operations.

The searchable interface allows additional, optional search operations. Currently, this specification defines only one optional operation. **XQuerySearch** returns selected information extracted from resource descriptions that match a set of constraints expressed using the XQuery syntax [XQuery].

These operations are implemented to accept input parameters and return output results as SOAP messages in compliance with the WSDL document given in Appendix A.1. Compliant searchable registries must return a copy of the WSDL document whenever a client invokes the service endpoint URL with the **"?wsdl"** argument appended to it. The returned WSDL may contain additional operations beyond those specified in this section; however, the original searching operations, their arguments and their outputs must not be altered.

## 2.1 Required Search Operations

The two search operations—**Search** and **KeywordSearch**—return resource records that match a set of selection criteria.

### 2.1.1 Output Format

The two search operations share a common output format for the resource records that match the search criteria. The response is a SOAP message in compliance with the WSDL document given in Appendix A.1. This message is defined to have a single part: a `SearchResponse` element from the http://www.ivoa.net/wsdl/RegistrySearch/v1.0 namespace. This element in turn wraps a single `VOResources` element from the http://www.ivoa.net/xml/RegistryInterface/v1.0 namespace (from now on referred to using the "`ri:`" prefix) that contains each of the matching records and conforms to the following XML Schema definition:

```
VOResources and Resource Element Definitions
<xs:element name="VOResources">
    <xs:complexType>
        <xs:sequence>
            <xs:choice>
                <xs:element ref="ri:Resource"
                            minOccurs="0" maxOccurs="unbounded"/>
                <xs:element name="identifier" type="vr:IdentifierURI"
                            minOccurs="0" maxOccurs="unbounded"/>
            </xs:choice>
        </xs:sequence>
        <xs:attribute name="from" type="xs:positiveInteger"
                      use="required" />
        <xs:attribute name="numberReturned" type="xs:positiveInteger"
                      use="required" />
        <xs:attribute name="more" type="xs:boolean" use="required" />
    </xs:complexType>
</xs:element>

<xs:element name="Resource" type="vr:Resource"/>
```

The required `ri:VOResources` attributes allow the results of the query to be "paged" over several calls that can be controlled via the operation input parameters, **from** and **max** (see sections 2.1.2 and 2.1.3). They assume that over some limited amount of time multiple calls to a search operation on a single registry with the same search constraints returns the same results and in the same order. If the client does not request paging via the **from** and **max** parameters, the service may choose to return partial results (setting the appropriate attributes) if the results exceed the service's own internal limits.

| VOResources Attributes | | |
|---|---|---|
| **Attribute** | **Definition** | |
| `from` | *Value Type:* | Integer |
| | *Semantic Meaning:* | the 1-relative position of the first record returned among the total set of matched elements. |
| | *Occurrences:* | required |
| `numberReturned` | *Value Type:* | Integer |
| | *Semantic Meaning:* | the number of records returned in this response. |
| | *Occurrences:* | required |
| `more` | *Value Type:* | Boolean |
| | *Semantic Meaning:* | If true, additional results are available beginning with a position of `from` + `numberReturned`. If false, no more results are available beyond this set. |
| | *Occurrences:* | required |

If all records matched by the query are returned in a single response, the value of **more** must be set to false and **from** must be set to 1.

```
Example:
<ri:VOResource from="100" numberReturned="200" more="true"
               xmlns:ri="http://www.ivoa.net/xml/RegistryInterface/v1.0"
...
</ri:VOResourced>
```

The contents of the `ri:VOResources` element depends on the value of the operation input parameter, **identifiersOnly**. If this parameter is set to true, then the `ri:VOResources` element must contain a list of `identifier` elements that contain the IVOA identifiers of resources that match the input query. If **identifiersOnly** is false, then the `ri:VOResources` element must contain a list of `ri:Resource` elements containing the full VOResource descriptions of resources that matches the query. Each child `ri:Resource` element is of type **Resource** from the VOResource XML Schema (having the namespace, http://www.ivoa.net/xml/VOResource/v1.0, from now on referred to using the "`vr:`" prefix), or a legal extension of the `vr:Resource` type. If the type of the

`Resource` element is actually an extension of the `vr:Resource` type, then the `Resource` element MUST specify the specific type using an `xsi:type` attribute (where the `xsi` prefix refers to the http://www.w3.org/2001/XMLSchema-instance namespace) in compliance with the XML Schema standard [Schema].

The search responses must include only active resource records—that is, records in which the `vr:Resource` element's `status` attribute is set to the value, "active". Thus, clients do not have to explicitly include a search constraint to avoid deleted or inactive records. (**GetResource** and **XQuerySearch** operations, on the other hand, may return inactive or deleted resources.)

The search responses must include the `xsi:schemaLocation` attribute (regardless of the value of **identifiersOnly**) in compliance with the XML Schema standard [Schema] to indicate a URL location for the VOResource schema and all of the legal extensions of VOResource that are employed in the response. This `xsi:schemaLocation` attribute must appear either as an attribute of the `ri:VOResources` element or as an attribute of each child `ri:Resource` element (when **identifiersOnly** is false) or both. When `xsi:schemaLocation` appears as an attribute of `ri:Resource`, locations need only be given for the schemas employed within that resource. The URL location for the VOResource core schema (http://www.ivoa.net/xml/VOResource/v1.0) must be set to "http://www.ivoa.net/xml/VOResource/v1.0". For those legal extensions that are standard schemas recognized by the IVOA, the location should be set to the standard location in the IVOA Document repository whose URL begins with "http://www.ivoa.net/xml/".

---

**Example:** This illustrates the use of `xsi:type` and `xsi:schemaLocation` attributes in the search output results. the `xsi:schemaLocation` attribute contains pairs of values where the first value is the schema namespace and the second value is the URL location of that schema. For IVOA standard schemas, the namespace can be used as the URL location.

```
<ri:VOResources
    xmlns:ri="http://www.ivoa.net/xml/RegistryInterface/v1.0"
    xmlns:vs="http://www.ivoa.net/xml/VODataService/v1.0"
    xmlns:vg="http://www.ivoa.net/xml/Registry/v1.0"
    xsi:schemaLocation="http://www.ivoa.net/xml/VOResource/v1.0
                        http://www.ivoa.net/xml/VOResource/v1.0
                        http://www.ivoa.net/xml/VODataService/v1.0
                        http://www.ivoa.net/xml/VODataService/v1.0
                        http://www.ivoa.net/xml/SIA/v1.0
                        http://www.ivoa.net/xml/SIA/v1.0
                        http://www.ivoa.net/xml/Registry/v1.0
                        http://www.ivoa.net/xml/Registry/v1.0">

    <ri:Resource xsi:type="vs:CatalogService" status="active" …>
    …
    </ri:Resource>

    <ri:Resource xsi:type="vs:Registry" status="active" …>
    …
    </ri:Resource>
</ri:VOResources>
```

If a legal search query does not match any resource records, the `ri:VOResources` element must contain no `ri:Resource` elements. If the input search query is illegal in its syntax or the operation encounters any other error that prevents returning the requested records, the operation must return an **ErrorResponse** fault, represented by an `ErrorResponse` element:

**ErrorResponse Element Definition**
```
<xs:element name="ErrorResponse">
    <xs:complexType>
        <xs:sequence>
            <xs:element name="errorMessage" type="xs:string"
                        minOccurs="1" maxOccurs="1" />
        </xs:sequence>
    </xs:complexType>
</xs:element>
```

An `ErrorResponse` element must include a human-oriented error message describing the nature of the error.

### 2.1.2 Constraint-based Search Query

The **Search** operation allows clients to retrieve a list of resource descriptions that match constraints of values corresponding to specific metadata from the VOResource schema (and its legal extensions). The operation's input message is defined to have a single part, a `Search` element, which contains four child elements that serve as the four input parameters:

---

**Search Element Definition**

```
<xs:element name="Search">
   <xs:complexType>
      <xs:sequence>
         <xs:element ref="tns:Where" minOccurs="1" maxOccurs="1" />
         <xs:element name="from" type="xs:positiveInteger"
                     minOccurs="0" maxOccurs="1" />
         <xs:element name="max" type="xs:positiveInteger"
                     minOccurs="0" maxOccurs="1" />
         <xs:element name="identifiersOnly" type="xs:boolean"
                     minOccurs="0" maxOccurs="1" />
      </xs:sequence>
   </xs:complexType>
</xs:element>
```

---

| Search Parameter Elements | | |
|---|---|---|
| **Element** | **Definition** | |
| `Where` | *Value Type:* | an ADQL/x Where clause |
| | *Semantic Meaning:* | the constraints to use for selecting matched resource records |
| | *Occurrences:* | required |
| `from` | *Value Type:* | integer |
| | *Semantic Meaning:* | the minimum position in the complete set of matched records of the range of records to be returned |
| | *Occurrences:* | optional; default: 1 |
| `max` | *Value Type:* | integer |
| | *Semantic Meaning:* | the maximum number of matched records to return. The service may choose to return fewer. |
| | *Occurrences:* | optional; default: the maximum that the service can deliver |
| `identifiersOnly` | *Value Type:* | boolean |
| | *Semantic Meaning:* | If true, return the results as a list of identifiers; if false, return as a list of complete resource descriptions. |
| | *Occurrences:* | optional; default: false |

The one required parameter, the `Where` element, is of type `whereType` from the ADQL XML Schema [Appendix A.5] (having the namespace, http://www.ivoa.net/xml/ADQL/v1.0, from now on referred to using the "`adql:`" prefix; see Appendix A.1) which contains the constraints that specific components of the resource metadata must satisfy.

```
Example:   a Search SOAP request message:

<soapenv:Envelope
          xmlns:soapenv="http://schemas.xmlsoap.org/soap/envelope/"
          xmlns:xsi="http://www.w3.org/2001/XMLSchema-instance">
  <soapenv:Body>

    <rs:Search xmlns="http://www.ivoa.net/wsdl/RegistrySearch/v1.0"
               xmlns:rs="http://www.ivoa.net/wsdl/RegistrySearch/v1.0"
               xmlns:adql="http://www.ivoa.net/xml/ADQL/v1.0">
      <rs:Where>
        <adql:Condition xsi:type="adql:likePredType">
          <adql:Arg Table="" xsi:type="adql:columnReferenceType"
                      name="description"
                      xpathName="content/description"/>
          <adql:Pattern xsi:type="adql:atomType">
            <adql:Literal Value="%quasar%" xsi:type="adql:stringType"/>
          </adql:Pattern>
          </adql:Condition>
        </rs:Where>
        <max xmlns="">500</max>
        <identifiersOnly xmlns="">false</identifiersOnly>
      </rs:Search>

  </soapenv:Body>
</soapenv:Envelope>
```

The specific components are named within search constraints (represented by `adql:Condition` elements) using `adql:Arg` elements subject to the following restrictions:

- The `Table` attribute, which is required by the ADQL Schema, should be set to an empty string and must be ignored by the **Search** method implementation.
- The `Name` attribute, which is required by the ADQL Schema, may be set to an empty string or to a short name to serve as an alias for the resource metadata referred to. This value must be ignored by the **Search** method.
- The `xpathName` attribute must be set to a restricted XPath string, subject to the rules in section 2.2.1. This XPath string identifies the specific VOResource element (or legal extension) within the resource record that is to be constrained.

The **Search** implementation must implicitly add the constraint that the `status` attribute on the `ri:Resource` element be equal to "active"; deleted and inactive records, therefore, should not be returned.

Matched resource records are encoded using the VOResource XML Schema (and its legal extensions) according to the specifications given in the Search WSDL and described in Section 2, and they should include all information available to the registry that is compliant with the VOResources definitions.

### 2.1.2.1     Restrictions on the use of XPath in ADQL

The value of the `xpathName` attribute in any `adql:Arg` element used within the input to the Search method must be a legal XPath [XPath] string that is restricted in form by the following rules:

- The path points to an element or attribute value within a resource description encoded with the VOResource schema and/or any of its legal extensions.
- When the path points to a specific element, that element must be of a simple type as defined by the XML Schema standard [Schema].
- The path is relative and assumes that the context node is the element that forms the parent of a single resource description (e.g. a `Resource` element) and is of type `vr:Resource` or one of its legal extensions.
- The path must be composed only of location steps with child axes expressed using the abbreviated syntax for child elements and attributes: elements are referred to simply by their name, and attributes are referred to by their name preceded by an '@' character.  Unabbreviated location steps—i.e., those that require the double colon ('::') syntax—are not allowed.  All other types of abbreviated axes, including use of double slashes ('//'), single and double periods ('.' and '..'), and wildcards ('*'), are not allowed.
- The path must not include any predicates (i.e., qualifiers expressed using square brackets, '[…]').
- When "xsi" is used as an attribute prefix, it is implicitly assumed to refer to the http://www.w3.org/2001/XMLSchema-instance namespace.

| Legal Examples: | |
|---|---|
| curation/publisher | the resource publisher's name |
| curation/publisher/@ivo-id | the publisher's IVOA identifier |
| @xsi:type | the specific type of resource |
| capability/interface/@xsi:type | the specific type of interface |
| **Illegal Examples:** | |
| Resource/title | wrong context node |
| content | not an element with a simple type |
| curation/child::publisher | "child::" syntax not allowed |
| curation//@ivo-id | "//" syntax not allowed |
| capability[@xsi-type="vg:Harvest"]/accessURL | "[…]" syntax not allowed |

This restricted form of XPath is intended to make it straight forward to transform the ADQL Where clause to a string-based query—namely SQL and XQuery—through a static mapping from an XPath to an attribute in a local database without parsing the internal content of the path.

> **Note:**
> Because VOResource schema and its legal extensions set `elementFormDefault` and `attributeFormDefault` to both be 'unqualified', prefixes are not normally required to qualify elements and attributes. xsi:type is an exception because it is technically a global attribute, and `attributeFormDefault` does not apply; thus, the prefix would be required in a standard XPath, which is why the last rule is needed. A known exception at this time is the case of the Space-Time Coordinates schema (STC, http://www.ivoa.net/xml/STC/stc-v1.30.xsd), which defines many global elements; thus, technically, these elements would require prefixes, too. This document does not address the problem of querying against these elements because it specifies no additional rules for understanding the mapping of other prefixes used in the context of an ADQL query. Because of the complexity of STC, it is not likely that normal ADQL (or XQuery) queries will be particularly useful. Thus, the problems of invoking other prefixes within ADQL and generally querying against STC are left to be solved in a future version of this document.

> **Note:**
> It is important to note that search operations do not support resource harvesting described in section 3. Normally, an end-user would use search to retrieve resource descriptions, but not to selectively harvest information between registries.

> **Note:**
> Because this specification does not provide any rules for understanding the prefixes that might appear in an XPath used in an ADQL where clause (apart from `xsi`), it is not obvious how best to query against the values of `xsi:type` attributes. Section 3.1.2 strongly recommends that publishing registries export VOResource records using the prefixes listed in Appendix 4 where appropriate. This allows a client to be explicit about a desired `xsi:type` value, as in this constraint expressed in ADQL/s:
>
> | | |
> |---|---|
> | capability/@xsi:type = 'cs:ConeSearch' | Select services supporting the ConeSearch capability. |
>
> For when a registry does not follow this recommendation (or a client does not wish to trust that it does), here are some recommended techniques for matching the `xsi:type`:
>
> | | |
> |---|---|
> | capability/@xsi:type like '%:ConeSearch' | Select services supporting the ConeSearch capability. |
> | @xsi:type like '%:Service' | Select generic Service resources only |
> | @xsi:type like '%Service' | Select any resource with "Service" in its name, including "Service," "DataService," and "CatalogService." |

### 2.1.3  Keyword Search Query

The purpose of the **KeywordSearch** operation is to provide a simple way to select resources based on the string values in their resource descriptions. The operation only queries for active resources noted by `status='active'`. The operation's input message is defined to have a single part, a `KeywordSearch` element, which contains five child elements that serve as the five input parameters:

```
KeywordSearch Element Definition
<xs:element name="KeywordSearch">
   <xs:complexType>
      <xs:sequence>
         <xs:element name="keywords" type="xs:string"
                     minOccurs="1" maxOccurs="1" />
         <xs:element name="orValues" type="xs:boolean"
                     minOccurs="0" maxOccurs="1" default="true"/>
         <xs:element name="from" type="xs:positiveInteger"
                     minOccurs="0" maxOccurs="1" />
         <xs:element name="max" type="xs:positiveInteger"
                     minOccurs="0" maxOccurs="1" />
         <xs:element name="identifiersOnly" type="xs:boolean"
                     minOccurs="0" maxOccurs="1" />
      </xs:sequence>
   </xs:complexType>
</xs:element>
```

The meaning of the last three parameters above and their effect on the output is the same as for the **Search** operation.

The `keywords` parameter is a string that consists of one or more words separated by whitespace characters.  The characters that qualify as whitespace are the same as in XML: space (x20), tab (x9), line feed (xA), and carriage return (xD).  A phrase is a portion of `keywords` parameter that is enclosed in double quotation marks (e.g. "black hole").  Words and phrases are extracted from the `keywords` parameter as a list of tokens to be used in the search.  When a phrase is extracted from the parameter, the quotation marks are removed.

The **KeywordSearch** implementation forms a query by, in effect, creating a search constraint for each word or phrase in this parameter.  For each active or inactive resource record, each word or phrase is compared against every value for a selected set of resource metadata that includes at minimum the following (drawn from the VOResource schema):

- `identifier`: the resource's IVOA identifier
- `content/description`: the descriptive summary of the resource
- `title`: the resource title

- **@xsi:type**:  the specific type of resource specified as an extension of the **Resource** type
- **content/subject**:  the subject topics associated with the resource
- **content/type**:  the general type of resource

| KeywordSearch Parameter Elements | | |
|---|---|---|
| **Element** | **Definition** | |
| `keywords` | *Value Type:* | text string |
| | *Semantic Meaning:* | the list of words or phrases to search for within resource descriptions |
| | *Occurrences:* | Required |
| `orValues` | *Value Type:* | Boolean |
| | *Semantic Meaning:* | if true, apply multiple word/phrase constraints with a logical OR; if false, apply with a logical AND |
| | *Occurrences:* | Required |
| `from` | *Value Type:* | Integer |
| | *Semantic Meaning:* | the minimum position in the complete set of matched records of the range of records to be returned |
| | *Occurrences:* | Optional; default: 1 |
| `max` | *Value Type:* | Integer |
| | *Semantic Meaning:* | the maximum number of matched records to return.  The service may choose to return fewer. |
| | *Occurrences:* | optional; default:  the maximum that the service can deliver |
| `identifiersOnly` | *Value Type:* | Boolean |
| | *Semantic Meaning:* | If true, return the results as a list of identifiers; if false, return as a list of complete resource descriptions. |
| | *Occurrences:* | Optional; default: false |

The implementer may include additional metadata values in the comparison as they choose (which may include non-string values). It is legal to compare the word with all simple type values in the record.  If the word or phrase is contained within one of the selected set of resource metadata values, the constraint evaluates as TRUE.   It is up to the implementer to decide what it means for a word to be considered "contained;" for example, the implementation may also test for related forms of the word.  It is also up to the implementer to determine how to match a phrase—in particular, how to match the separation between words (e.g. whether spaces are strictly matched). The implementer may further parse the phrase into words and include comparison constraints on those individual words.

The results of all of the constraint tests (one for each word) are combined logically according to the value of **orValues**: if **orValues** is TRUE, then the resource record is returned when any of the constraints are TRUE, and if it FALSE, then all constraints must be TRUE in order for the record to be returned.

Matched resource records are then encoded using the VOResource XML Schema (and its legal extensions) and should include all information available to the registry that complies with the VOResource standard.

> **Note:**
> This specification provides wide latitude in how the **KeywordSearch** is implemented; thus, different registries may return different results to the same set of input keywords. If precision and consistency in results is important regardless of which registry is queried, users should favor the Search or XQuerySearch operations.

## 2.2  Resolve Operations

The two resolve operations—**GetResource** and **GetIdentity**—each select and return a single resource record.

### 2.2.1  Output Format

The two resolve operations share a common output format for returning a single resource record. The response is a SOAP message in compliance with the WSDL document given in Appendix A.1. This message is defined to have a single part: a `ResolveResponse` element from the http://www.ivoa.net/wsdl/RegistrySearch/v1.0 namespace. This element in turn wraps an `ri:Resource` element of type `vr:Resource` or one of its legal extensions. As in with the search operations when the type of the `ri:Resource` element is actually an extension of the `vr:Resource` type, then the `ri:Resource` element MUST specify the specific type using an `xsi:type` attribute.

The `Resource` element must include a `xsi:schemaLocation` attribute in compliance with the XML Schema standard [Schema] to indicate a URL location for the VOResource schema and all of the legal extensions of VOResource that are employed in the response. As with the search operation responses, the URL location for the VOResource core schema (http://www.ivoa.net/xml/VOResource/v1.0) must be set to "http://www.ivoa.net/xml/VOResource/v1.0". For those legal extensions that are standard schemas recognized by the IVOA, the location should be set to the standard location in the IVOA Document repository whose URL begins with "http://www.ivoa.net/xml/".

### 2.2.2 Identifier Resolution

The purpose of the **GetResource** operation is to provide a simple way to resolve a unique IVOA Identifier to a full resource description. The input message is defined to have a single part, a `GetResource` element. This element contains the operation's one input parameter, `identifier`, of type `vr:IdentifierURI`, encodes an IVOA identifier. The output message contains a single VOResource record whose `identifier` element matches the input identifier.

If the registry does not have a resource record (or otherwise cannot access one) with an identifier matching the input parameter, the **GetResource** operation should return a **NotFound** fault message, represented by a `NotFound` element:

```
NotFound Element Definition
<xs:element name="NotFound">
    <xs:complexType>
        <xs:sequence>
            <xs:element name="errorMessage" type="xs:string"
                        minOccurs="0" maxOccurs="1" />
        </xs:sequence>
    </xs:complexType>
</xs:element>
```

Including an error message in the fault response is optional but recommended.

If the operation encounters any error that prevents it from determining whether the identifier can be resolved to a description or otherwise prevents the delivery of that description, the operation must return an **ErrorResponse** fault as described in section 2.1.1.

### 2.2.2   Identity Query

The purpose of the **GetIdentity** operation is to provide a simple way to get the VOResource record that describes the implementing registry itself. A client may then inspect this VOResource record to discover various information about the implemented registry (See section 2.7).

The **GetIdentity** operation takes no parameters. The result is a single VOResource record whose format conforms to the format described in section 2.2.1; however, with this operation, the `ri: Resource` element must include an `xsi:type` attribute set to indicate the `Registry` resource extension type from the VORegistry extension schema (having the namespace http://www.ivoa.net/xml/VORegistry/v1.0, from now on referred to using the "`vg:`" prefix). The recommended value to express this type is "`vg:Registry`". The VORegistry schema is described in section 4; see Appendix A.3 for the full XML Schema definition.

## 2.3  XQuery Search

**XQuerySearch** is an optional operation of the searching interface that allows clients to form constraint-based queries with greater control than the required **Search** operation.  It also allows the client to control the format of the query output; in particular, the client can obtain only the metadata needed rather than the full Resource record.  The client can determine if a searchable registry supports this operation by consulting the registry's resource description (see Section 4).

The operation's input message is defined to have a single part, an **XQuerySearch** element.  This element contains the operation's one input string parameter, **xquery**:

---

**XQuerySearch Element Definition**
```
<xs:element name="XQuerySearch">
   <xs:complexType>
      <xs:sequence>
         <xs:element name="xquery" type="xs:string"
                     minOccurs="1" maxOccurs="1" />
      </xs:sequence>
   </xs:complexType>
</xs:element>
```

---

The value of the **xquery** element is a string that states the query and must conform to the XQuery syntax [XQuery].  The XPath strings [XPath] used in the query must be written as if each resource record is stored as a separate document under a root element called **RootResource**.  The operation implementation may translate the XPath as necessary to reflect the actual storage of the records within the registry.

The operation's output message is also defined to have a single part, an **XQuerySearchResponse** element having the following definition:

---

**XQuerySearchResponse Element Definition**
```
<xs:element name="XQuerySearchResponse">
   <xs:complexType>
      <xs:sequence>
         <xs:any minOccurs="0" maxOccurs="1" />
      </xs:sequence>
   </xs:complexType>
</xs:element>
```

---

The specific XML content of the `XQuerySearchResponse` must comply with the format requested by the XQuery input query.

If a registry does not support XQuery–based queries, the **XQuerySearch** operation must be implemented to always return an **UnsupportedOperation** fault message. This message has a single part in the form of an `UnsupportedOperation` element:

```
UnsupportedOperation Element Definition
<xs:element name="UnsupportedOperation">
   <xs:complexType>
      <xs:sequence>
         <xs:element name="errorMessage" type="xs:string"
                     minOccurs="0" maxOccurs="1" />
      </xs:sequence>
   </xs:complexType>
</xs:element>
```

Including an error message in the fault response is optional but recommended.

All other error conditions encountered by the implementation that prevent the return of the query results should be handled by returning an **ErrorResponse** fault as described in section 2.1.1.

# 3   Harvesting

**Harvesting** is the mechanism by which a registry can collect resource descriptions from other registries. This mechanism is used by full searchable registries to aggregate resource descriptions from many publishing registries. It can also be used to synchronize two registries to ensure that they have the same contents. This section defines the **IVOA Harvesting Interface**. Client applications that make use of this interface are referred to as **harvesters.** Those registries that declare themselves as harvestable (section 4) must comply with the specification described in this section.

## 3.1 Harvesting Interface

This specification defines two variants of the harvesting interface, both built on the standard Protocol for Metadata Harvesting developed by Open Archives Initiative (OAI-PMH) [OAI]. The first variant is one that is fully compliant with the OAI-PMH version 2.0 standard; harvestable registries must support this variant. Compliance with this base standard allows IVOA registries to be accessed by applications from outside the IVOA community. The second variant, a Web Service version of the OAI interface (in which the input and outputs are transported as SOAP messages), is also defined as an optional alternative. In addition to basic OAI-PMH compliance, this specification defines an additional set of OAI-PMH-compliant requirements and recommendations that are described in sections 3.1.1 through 3.1.6 below.  These apply to both variants of the harvesting interface.

### 3.1.1   A Summary of the OAI Interface

The required variant of the interface is defined by the OAI-PMH v2.0 specification [OAI], which itself defines:

- the meaning and behavior of the six harvesting operations, referred to as "verbs",
- the meaning of the input arguments for each operation, and
- the XML Schema used to encode response messages.

The optional Web Service variant of the interface maps the meaning, behavior, and schema of the OAI-PMH specification into a Web Service Definition Language (WSDL) document (see Appendix A.2); this WSDL defines:

- the six "verbs" defined as Web Service operations
- SOAP encoding of the operation input arguments and response messages, based on the OAI-PMH XML Schema.

In summary, the OAI-PMH standard defines six operations:

**Identify**:  provides a description of the registry
**ListIdentifiers**:  returns a list of identifiers for the resource records held by the registry.
**ListRecords**:  returns all Resource records in the registry.  Registries may use the set "ivo_managed" to get Resource records managed by this particular registry.
**GetRecord**: returns a single resource description matching a given identifier.
**ListMetadataFormats**: returns a list of supported formats that the registry can use to encode resource descriptions upon a harvester's request.
**ListSets**:  return a list of category names supported by the registry that harvesters can request in order to get back a subset of the descriptions held by the registry.

The **ListRecords** and **GetRecord** operations return the actual resource description records held by the registry.  These descriptions are encoded in XML and wrapped in a general-purpose envelope defined by the OAI-PMH XML Schema (with the namespace http://www.openarchives.org/OAI/2.0).

Through the operations' arguments, OAI-PMH provides a number of useful features:

- *Support for multiple return formats*.  As suggested by the **ListMetadataFormats** operation, a harvester can request the formats available for encoding returned resource descriptions.
- *Harvesting by date*.  The **ListIdentifiers** and **ListRecords** operations both support "from" and "until" date arguments.  The "from" argument can be used to retrieve records that have changed since the last harvest.
- *Harvesting by category*.  The **ListIdentifiers** and **ListRecords** operations both support a "set" argument for retrieving resources that are grouped in a particular category.  Resource records may belong to multiple groups.
- *Marking records as deleted*.  Registries may mark records as deleted so that harvesters may remove access to them from their applications.  Registries may permanently remove deleted resources that have been marked deleted for more than six months.
- *Support for resumption tokens*.  If a request results in returning a very large number of records, the registry can choose to split the results over several calls; this is done by passing a resumption token back to the harvester.  The harvester uses it to retrieve the next set of matching results.
- *Harvesting with no date*.  Deleted resource records may not be returned when no "from" or "until" is specified.

> **Note:**
> The Web Service version of the OAI-PMH protocol has been designed to match the behavior and functionality of the original version as much as possible.  One reason for this is to make it as straightforward as possible to build bridges between implementations of both types and to build off the existing OAI software.

> **Note:**
> It is important to note that the OAI-PMH interface is not intended to be a general search interface.  The filtering capabilities described above are just enough to support intelligent harvesting between registries.  Most end-user applications will use the search interface described in sections 3 and 4 to retrieve resource descriptions.

The Web Service or SOAP version of OAI-PMH augments the original specification with a standard Web Service Definition Language (WSDL) document, which is listed in Appendix A.2.  Harvestable registries complying to the SOAP version of OAI-PMH must emit a copy of the WSDL document, with a service element appropriate for the local endpoint URL added in, in response to a call to the Web Service URL with the standard "?wsdl" argument.  All six of the standard operations must be implemented.  Additional, non-standard operations may be added; however, the definition of the six standard operations, along with the definition of their inputs and outputs, must not be altered.  The interface is recognized as the OAI-PMH standard when the default namespace for the WSDL matches "http://www.ivoa.net/wsdl/RegistryHarvest/v1.0" exactly.

The subsequent sections below describe how the standard OAI-PMH features are used to support IVOA-specific functionality.

### 3.1.2 Metadata Formats for Resource Descriptions

All IVOA registries that support the Harvesting Interface must support two standard metadata formats: the OAI Dublin Core format (mandated by the base OAI-PMH standard) and the IVOA VOResource metadata format [VOResource].

The VOResource metadata format will have the metadata prefix name "ivo_vor" which can be used wherever an OAI-PMH metadata prefix name is supported (see OAI standard, section 3.4, "metadataPrefix and Metadata Schema"). The format uses the VOResource core XML Schema with the namespace http://www.ivoa.net/xml/VOResource/v1.0 (referred hereto with the namespace prefix "`vr`") along with any legal extension of this schema to encode the resource descriptions within the OAI-PMH `metadata` tag from the OAI XML Schema (namespace http://www.openarchives.org/OAI/2.0, hereto referred by the namespace prefix "`oai`"). The format is specifically represented by an element called `Resource` from the http://www.ivoa.net/xml/RegistryInterface/v1.0 namespace (from now on referred to using the "`ri:`" prefix) as the sole child of the `oai:metadata` element. The `ri:Resource` element must include an `xsi:type` attribute that assigns the element's type to `vr:Resource` or one of its legal extensions.

> **Note:**
> If and when the VOResource schema evolves to a new version, this standard must be updated accordingly. Thus, this definition is locked to particular version of the VOResource, so saying that a registry is compliant with vX.X of this document implies a specific version of VOResource.

It is strongly recommended that all QName values of `xsi:type` attributes within the VOResource record use namespace prefixes drawn from the recommended list given in Appendix A.4 when appropriate. When the type is drawn from an IVOA standard extension schema not listed in A.4, the prefix recommended by the standard itself should be used. It is also strongly recommended these prefixes be used for storage in the registry to facilitate easier searching in the search interface (see Note at the end of section 2.1.2).

The OAI Dublin Core format, with the metadata prefix of "oai_dc", is defined by the OAI-PMH base standard and must be supported by all OAI-PMH compliant registries. This document does not specify how a record in the VOResource format maps into the OAI Dublin Core format; however, the IVOA Registry Working Group may recommend such a mapping based on the IVOA Resource Metadata standard.

Harvestable registries may support other metadata formats. The **ListMetadataFormats** must list all names for formats supported by the registry; this list must include "ivo_vor" and "oai_dc".

### 3.1.3   Identifiers in OAI Messages

In accordance with the OAI-PMH standard, an OAI-PMH XML envelope that contains a resource description must include a globally unique URI that identifies that resource record. This identifier must be the IVOA identifier used to identify the resource being described and cited as the value of the `vr:identifier` resource metadata.

> **Note:**
> This specification does not follow the recommendation of the OAI-PMH standard with regard to record identifiers. OAI-PMH makes a distinction between the resource record containing resource metadata and the resource itself; thus, it recommends that the identifier in the OAI envelope be different from the resource identifier. In particular, the former is the choice of the publishing registry. This allows one to distinguish resource descriptions of the same resource from different registries, which in principle could be different.
>
> In the VO, because it is intended that resource descriptions of the same resource from different registries should not differ (apart from their `validationLevel` [VOResource]), there is not a strong need to distinguish between the resource and the resource description. By making the resource and resource record identifiers the same, it makes it much easier to retrieve the record for a single resource via **GetRecord**, regardless of which registry is being queried. Otherwise—when the registry chooses the record identifier—a client will not *a priori* know the record identifier for a particular resource, and so it is left to call **ListRecords** and search through the metadata of all the records itself to find the one of interest. In contrast, IVOA identifiers are intended to be a cross-application way of referring to a resource, and thus when a client wants only a single specific resource record, it is very likely that it would know the resource identifier when making a call to the **GetRecord** operation.

### 3.1.4   Required Records

This section describes the records that a harvestable IVOA Registry must include among those it emits via the OAI-PMH operations.

The harvestable registry must return one record that describes the registry itself as a whole, and the "ivo_vor" format must be supported for this record. This record is included in the **Identify** operation response (see section 3.1.5). When encoded using the "ivo_vor" format, the returned `ri:Resource` element must be of the type `Registry` from the VORegistry schema (namespace http://www.ivoa.net/xml/VORegistry/v1.0; hereto referred by the "vg" namespace prefix). The record must include a `vg:managedAuthority` for every Authority Identifier [Identifiers] that originated at that registry. The registry may contain other registry records for other registries it knows about; use of a `vr:Resource` extension type other than `vg:Registry` to describe these other registries is strongly discouraged.

The harvestable registry must return exactly one record in "ivo_vor" format for each Authority Identifier listed as a `vg:managedAuthority` in the `vg:Registry` record that describes that registry. When encoded in the "ivo_vor" format, the type of `ri:Resource` must be `vg:Authority`.

### 3.1.5 The Identify Operation

The **Identify** operation describes the harvestable registry as a whole. The response from this operation must include all information required by the OAI-PMH standard. In particular, it must include an `oai:baseURL` element that must refer to the base URL to the harvesting interface endpoint. When the **Identify** operation is called through the Web Service variant, the `oai:baseURL` element value must be the endpoint of the Web Service itself (i.e. the URL used to retrieve the WSDL document via the standard URL suffix, "?wsdl").

The **Identify** response must include an `oai:description` element containing a single `Resource` element with an `xsi:type` attribute that sets the element's type to `vg:Registry`. The content of `vg:Registry` type must be the registry description of the harvestable registry itself.

### 3.1.6 IVOA Supported Sets

**Sets**, as defined in the OAI-PMH standard, "[are] an optional construct for grouping items for the purpose of selective harvesting" (see the OAI-PMH standard, section 2.6). Harvestable IVOA registries are free to define any number of custom sets for categorizing records. The OAI-PMH standard allows a record to be a member of multiple sets. This specification defines one reserved set name with a special meaning; future versions of this specification may define additional set names. These reserved set names will all start with the characters "ivo_"; implementers should not define their own set names that begin with this string. While support for sets is optional to be compliant with the OAI-PMH standard, a harvestable registry must support the set with the reserved name "ivo_managed" to be compliant with this specification.

The "ivo_managed" set refers to all records that originate from the queried registry. That is, those records that were harvested from other registries are excluded. The IVOA Resource identifiers given in the records must have an Authority Identifier that matches on one of the `vg:managedAuthority` values in the `vg:Registry` record for that registry. Full searchable registries may use this set to avoid getting duplicate records when harvesting from many registries.

All sets that are supported by the harvestable registry, including the one required set, must be listed in the response to the **ListSets** operation in compliance with the OAI-PMH standard.

## 3.2 Harvesters

A registry that collects resource descriptions from other registries through the Harvesting Interface defined above in section 3.1 is referred to as a **harvester.** A full registry attempts to establish a complete collection of all resource descriptions known to the VO either by replicating the contents of another full registry, or—more commonly—by selectively harvesting from all known publishing registries. Typically in the latter case, the harvester periodically engages the **ListRecords** operation of each know publishing registry with the **metadataPrefix** parameter set to "ivo_vor", the **set** parameter set to "ivo_managed", and the **from** parameter set to the time of the last successful harvest for that publishing registry.

Any registry that claims to be a full registry (see `vg:full` metadata defined in Section 4) must accept all records it harvests that are compliant with the VOResource metadata standard [VOResource], even if the resource type is not one that is recognized by the registry. Whenever any registry (full or not) exports a harvested record—through either the searching or the harvesting interface—it must return the complete record in its original format. The only change in the informational content allowed between harvesting and subsequent export is in the addition or removal of `vr:validationLevel` elements; more specifically, the registry may remove any or all of the `vr:validationLevel` elements in the record received via harvesting, and it may add `vr:validationLevel` elements in compliance with VOResource metadata standard and with a `validatedBy` attribute set to the registry's IVOA Identifier.

> **Note:**
> It is not intended that "original format" to mean a byte-for-byte copy; rather, it means that the descriptions are equivalent (apart from the **vr:validationLevel** elements) in an XML sense after discarding all ignorable whitespace.

> **Note:**
> The **vr:validationLevel** element provides a mechanism for a registry to rate the quality of a resource description and it's adherence to relevant standards. Its usage is covered in the VOResource standard [VOResource].

In the case of registry replication, the harvester can harvest from just the registry it is trying to replicate; the **ListRecords** operation can be used in much the same way except that the **set** parameter is not provided in order to get all records from that registry.

> **Note:**
> This document does not specify how a registry obtains the complete list of publishing registries, nor does it specify how a new publishing registry should make itself known to harvesters as both these issues are considered outside the scope of this specification.

As of this writing, the IVOA Registry Working Group has established a mechanism for discovering publishing registries in the form of a so-called, "Registry of Registries" [RofR]. Hosted by IVOA, it provides a browser-based interface for registering a publishing registry. The RofR implements the harvesting interface; thus harvesters can regularly consult this registry via OAI-PMH to retrieve the vg:Registry records of available harvestees.

Harvesters can determine how to harvest from a registry by consulting its VOResource description. Section 4 describes the VOResource extension used to describe a registry and the interfaces it supports along with an example.

# 4   Registering Registries

This specification defines a VOResource extension schema called VORegistry that can be used to specifically describe a registry and its support for the registry interface described in this document. These descriptions can be stored as resource records in registries. The schema is also used to register a *naming authority*—a publisher who claims ownership of an *authority identifier* from which IVOA identifiers may be created [Identifiers]. A publishing registry is said to exclusively *manage* a naming authority on behalf of the owning publisher; this means that only that registry may publish records with IVOA identifiers using that authority identifier.

The full VORegistry syntax definition expressed in XML Schema is listed in Appendix A.3.

## 4.1  The Schema Namespace and Location

The VORegistry schema namespace is "http://www.ivoa.net/xml/VORegistry/v1.0". As with the core VOResource Schema, the namespace URI has been chosen to allow it to be resolved as a URL to the XML Schema document that defines the VORegistry schema. Applications may assume that the namespace URI is so resolvable. In particular, it is recommended the namespace URI be given as the location for the VORegistry schema within the `xsi:schemaLocation` attribute.

The namespace prefix, `vg`, is used by convention to represent the VORegistry schema. Registries and other applications are encouraged to follow this convention.

## 4.2  The Authority Resource Extension and the Publishing Process

The **vg:Authority** type extends the core **vr:Resource** type to specifically describe the ownership of an authority identifier [Identifiers] by a publishing organisation.

---

**vg:Authority Type Schema Definition**
```
<xs:complexType name="Authority">
   <xs:complexContent>
      <xs:extension base="vr:Resource">
         <xs:sequence>
            <xs:element name="managingOrg" type="vr:ResourceName"/>
         </xs:sequence>
      </xs:extension>
   </xs:complexContent>
</xs:complexType>
```

---

The IVOA identifier of a **vg:Authority** record provided via the **vr:identifier** element must have an empty resource key component [Identifiers]. The authority identifier component of the record's identifier is the one that is the subject of the record itself.

The **vg:Authority** type adds only one required item beyond the core VOResource metadata:

| vg:Registry Extension Elements | | |
|---|---|---|
| **Element** | **Definition** | |
| **managingOrg** | *Value Type:* | string with optional ID attribute: **vr:ResourceName** |
| | *Semantic Meaning:* | the organisation that owns or manages this authority identifier |
| | *Occurrences:* | required |
| | *Comments:* | This is almost always the organisation listed as the publisher of this Authority. |

The meaning of a **vg:Authority** record is that the organisation referenced in the **vg:managingOrg** element has the sole right to create (in collaboration with a publishing registry) and register resource descriptions using the authority identifier given by the **vr:identifier** element.

Before a publisher can create resource descriptions using a new authority identifier, it must first register its claim to the authority identifier by creating a **vg:Authority** record. Before the publishing registry commits the record for export, it must first search a full registry to determine if a **vg:Authority** with this identifier already exists; if it does, the publishing of the new **vg:Authority** record must fail. When a registry creates a **vg:Authority** record, it is said that the registry *manages* the associated authority identifier (on behalf of the owning publisher) because only that registry may create records with identifiers using that authority identifier.

## 4.3 Describing Registries with the Registry Resource Extension

The `vg:Registry` type extends the core `vr:Service` type to specifically describe registries that are compliant with this standard.

---

**vg:Registry Type Schema Definition**
```
<xs:complexType name="Registry">
   <xs:complexContent>
      <xs:extension base="vr:Service">
         <xs:sequence>
            <xs:element name="full" type="xs:boolean"/>
            <xs:element name="managedAuthority" type="vr:AuthorityID"
                        minOccurs="0" maxOccurs="unbounded"/>
         </xs:sequence>
      </xs:extension>
   </xs:complexContent>
</xs:complexType>
```

---

| vg:Registry Extension Metadata Elements | | |
|---|---|---|
| **Element** | **Definition** | |
| `Full` | *Value Type:* | boolean |
| | *Semantic Meaning:* | If true, this registry attempts to collect all resource records known to the IVOA |
| | *Occurrences:* | required |
| `managedAuthority` | *Value Type:* | an authority identifier: `vr:AuthorityID` |
| | *Semantic Meaning:* | an authority identifier that is managed by the registry |
| | *Occurrences:* | optional; multiple occurrences allowed |

If the `vg:full` element is set to true, the registry is obligated to accept all valid resource records it harvests from other registries in accordance with Section 4 of this specification.

The `vg:managedAuthority` element applies specifically to registries in their role as publishers of records. When a publishing registry claims to manage an authority identifier [Identifiers], it has created a `vg:Authority` resource record for that authority identifier (see section 4.2).

As a subclass of **vr:Service**, the **vg:Registry** type uses **vr:capability** elements to describe its support for the interfaces described in this specification. In particular, the VORegistry schema defines two extensions of the VOResource's **vr:Capability** type—one to describe the support for the searching interface and one to describe the harvesting interface—according to the recommendations for extension in the VOResource standard [VOResource, section 2.3.2]. Both extension types extension types extend from an intermediate restriction on **vr:Capability** called **vg:RegCapRestriction** to force the value of the standardID attribute to be "ivo://ivoa.net/std/Registry":

---

**vg:RegCapRestriction Type Schema Definition**

```
<xs:complexType name="RegCapRestriction" abstract="true">
   <xs:complexContent>
      <xs:restriction base="vg:Capability">
         <xs:sequence>
            <xs:element name="validationLevel" type="vr:Validation"
                        minOccurs="0" maxOccurs="unbounded"/>
            <xs:element name="description" type="xs:token"
                        minOccurs="0"/>
            <xs:element name="interface" type="vr:Interface"
                        minOccurs="0" maxOccurs="unbounded"/>
         </xs:sequence>
         <xs:attribute name="standardID" type="vr:IdentifierURI"
                       use="required"
                       fixed="ivo://ivoa.net/std/Registry"/>
      </xs:restriction>
   </xs:complexContent>
</xs:complexType>
```

---

As an abstract type, the **vg:RegCapRestriction** type cannot be used directly on its own within a resource description; one of the non-abstract extensions of this intermediate type must be used instead.

The **vr:Capability** extension types are invoked by applying the **xsi:type** attribute to the **vr:capability** element [VOResource, section 2.2.2]. If a registry supports both the searching and harvesting interfaces, the **vg:Registry** record should contain at least two **vr:capability** elements, one for each interface.

### 4.3.1  The Searching Capability

A registry declares itself to be a searchable registry by including a **vr:capability** element with an **xsi:type** attribute set to **vg:Search**.

---

**vg:Search Type Schema Definition**

```
<xs:complexType name="Search">
   <xs:complexContent>
```

---

```
      <xs:extension base="vg:RegCapRestriction">
         <xs:sequence>
            <xs:element name="maxRecords" type="xs:int"/>
            <xs:element name="extensionSearchSupport"
                        type="vg:ExtensionSearchSupport">
            <xs:element name="optionalProtocol"
                        type="vg:OptionalProtocol"
                        minOccurs="0" maxOccurs="unbounded">
         </xs:sequence>
      </xs:extension>
   </xs:complexContent>
</xs:complexType>
```

| vg:Search Capability Metadata Elements | |
| --- | --- |
| **Element** | **Definition** |
| `maxRecords` | *Value Type:* integer: `xsd:int` |
| | *Semantic Meaning:* The largest number of records that the registry search method will return. A value of zero or less indicates that there is no explicit limit. |
| | *Occurrences:* required |
| `extensionSearchSupport` | *Value Type:* String with controlled vocabulary |
| | *Semantic Meaning:* the level of support provided for searching against metadata defined in a legal VOResource extension schema. |
| | *Occurrences:* required |
| | `core` Only searches against the core VOResource metadata are supported. |
| | `partial` Searches against some VOResource extension metadata are supported but not necessarily all that exist in the registry. |
| | `full` Searches against all VOResource extension metadata contained in the registry are supported. |
| vg:Search Capability Metadata Elements (con't) | |
| **Element** | **Definition** |

| `optionalProtocol` | *Value Type:* | string with controlled vocabulary |
| | *Semantic Meaning:* | the name of an optional advanced search protocol supported. |
| | *Occurrences:* | optional |
| | *Allowed Values:* | `XQuery`  the XQuery  protocol as defined in section 2.3 |

A `vr:capability` element of type `vg:Search` must include at least one `vr:interface` element with an `xsi:type` attribute set to `vg:WebService` and the role attribute set to "std".  If the `vr:capability` element is used to simultaneously describe support for other versions of this Registry Interface standard, then the `vr:interface` element describing support for this version must include the `version` attribute set to "1.0".  The `vr:accessURL` element must be set to the endpoint URL for the Web Service interface that complies with section 2 of this specification.

> **Note:**
> The requirement that a vg:WebService interface appear within a vg:Registry record not enforced by the XML Schema document.  This requirement necessitates additional validation as described in the preface to the VOResource standard.

### 4.3.2  The Harvesting Capability

A registry declares itself to be a harvestable registry by including a `vr:capability` element with an `xsi:type` attribute set to `vg:Harvest`.

**vg:Harvest Type Schema Definition**
```
<xs:complexType name="Harvest">
    <xs:complexContent>
        <xs:extension base="vg:RegCapRestriction">
            <xs:sequence>
                <xs:element name="maxRecords" type="xs:int"/>
            </xs:sequence>
        </xs:extension>
    </xs:complexContent>
</xs:complexType>
```

**vg:Harvest Capability Metadata Elements**

| Element | Definition |
| --- | --- |

| `maxRecords` | Value Type: | integer: `xsd:int` |
| --- | --- | --- |
| | Semantic Meaning: | The largest number of records returned. A value greater than one implies that an OAI continuation token will be provided when the limit is reached. A value of zero or less indicates that there is no explicit limit and thus, continuation tokens are not supported. |
| | Occurrences: | Required |

The VORegistry schema defines two special extensions of the **`vr:Interface`** type that are used to indicate support for the OAI-PMH interface:

**vr:Interface Extension Types for OAI-PMH: Schema Definition**

```
<xs:complexType name="OAIHTTP">
    <xs:complexContent>
        <xs:extension base="vr:Interface">
            <xs:sequence/>
        </xs:extension>
    </xs:complexContent>
</xs:complexType>
<xs:complexType name="OAISOAP">
    <xs:complexContent>
        <xs:extension base="vr:WebService">
            <xs:sequence/>
        </xs:extension>
    </xs:complexContent>
</xs:complexType>
```

A **`vr:capability`** element of type **`vg:Harvest`** must include at least one **`vr:interface`** element with an **`xsi:type`** attribute set to **`vg:OAIHTTP`** and the **`role`** attribute set to "std". If the **`vr:capability`** element is used to simultaneously describe support for other versions of this Registry Interface standard, then the **`vr:interface`** element describing support for this version must include the **`version`** attribute set to "1.0". The **`vr:accessURL`** element must be set to the base URL for the OAI-PMH interface that complies with section 3 of this specification.

If the SOAP web service variant of OAI-PMH is supported, the record should include an additional **`vr:interface`** element with its type set to **`vg:OAISOAP`** and the **`role`** attribute set to "std:SOAP". If other versions of the SOAP harvesting interface are described in this same URL, the **`version`** attribute for OAI-SOAP interface must b set to "1.0".

**Registry Sample Instance Document**

```
<ri:Resource xsi:type="vg:Registry" xmlns=""
       xmlns:ri="http://www.ivoa.net/xml/RegistryInterface/v1.0"
       xmlns:vr="http://www.ivoa.net/xml/VOResource/v1.0"
       xmlns:vg="http://www.ivoa.net/xml/VORegistry/v1.0"
       xmlns:xsi="http://www.w3.org/2001/XMLSchema-instance"
       xsi:schemaLocation="http://www.ivoa.net/xml/VOResource/v1.0
                           http://www.ivoa.net/xml/VOResource/v1.0
                           http://www.ivoa.net/xml/VORegistry/v1.0
                           http://www.ivoa.net/xml/VORegistry/v1.0
                           http://www.ivoa.net/xml/RegistryInterface/v1.0
                           http://www.ivoa.net/xml/RegistryInterface/v1.0">
    <title>IVOA Registry of Registries sample entry</title>
    <shortName>RofR</shortName>
    <identifier>ivo://ivoa/registry</identifier>
    <curation>
      <publisher>
         IVOA
      </publisher>
      <creator>
        <name>Ray Plante</name>
      </creator>
      <date>2006-08-08</date>
      <contact>
        <name>Ray Plante</name>
        <email>rplante@ncsa.uiuc.edu</email>
      </contact>
    </curation>
    <content>
      <subject>registry repositories</subject>
      <description>
         This registry provides information regarding other registries.
      </description>
      <referenceURL>http://www.ivoa.net</referenceURL>
      <type>Registry</type>
      <contentLevel>Research</contentLevel>
    </content>
    <capability xsi:type="vg:Harvest"
                standardID="ivo://ivoa.net/std/Registry">
      <interface xsi:type="vg:OAIHTTPGet" role="std">
        <accessURL>
           http://www.ivoa.net/cgi-bin/rofr/oai.pl
        </accessURL>
      </interface>
      <interface xsi:type="vg:OAISOAP" role="std">
         <accessURL>
           http://www.ivoa.net/rofr/RegistryHarvest
         </accessURL>
      </interface>
      <maxRecords>100</maxRecords>
    </capability>

    <capability xsi:type="vg:Search"
                standardID="ivo://ivoa.net/std/Registry">

      <interface xsi:type="vr:WebService" role="std">
        <accessURL>
```

```
                http://nvo.ncsa.uiuc.edu/cgi-bin/nvo/search.pl
            </accessURL>
        </interface>
        <optionalProtocol>XQuery</optionalProtocol>
        <maxRecords>0</maxRecords>
    </capability>

    <full>false</full>
    <managedAuthority>ivoa</managedAuthority>
    <managedAuthority>ivoa.net</managedAuthority>
</ri:Resource>
```

# Appendix A.1  WSDL Document for Search Interface

Both WSDL documents for the search and harvest interfaces import the Registry
Interface schema that contains some common definitions; see the schema listing
after the WSDL listing below.

**RegistrySearch Interface WSDL**
```
<?xml version="1.0" encoding="UTF-8"?>
<definitions name="IVOARegistrySearch"
            xmlns="http://schemas.xmlsoap.org/wsdl/"
            xmlns:soap="http://schemas.xmlsoap.org/wsdl/soap/"
            xmlns:xs="http://www.w3.org/2001/XMLSchema"
            xmlns:adql="http://www.ivoa.net/xml/ADQL/v1.0"
            xmlns:ri="http://www.ivoa.net/xml/RegistryInterface/v1.0"
            xmlns:tns="http://www.ivoa.net/wsdl/RegistrySearch/v1.0
            targetNamespace="http://www.ivoa.net/wsdl/RegistrySearch/v1.0">
    <types>
        <xs:schema xmlns="http://www.w3.org/2001/XMLSchema"
                xmlns:tns="http://www.ivoa.net/wsdl/RegistrySearch/v1.0"
            targetNamespace="http://www.ivoa.net/wsdl/RegistrySearch/v1.0">

            <xs:import namespace="http://www.ivoa.net/xml/RegistryInterface/v1.0"
                schemaLocation="http://www.ivoa.net/xml/RegistryInterface/v1.0"/>
            <xs:import namespace="http://www.ivoa.net/xml/ADQL/v1.0"
                    schemaLocation="http://www.ivoa.net/xml/ADQL/v1.0" />

            <xs:element name="Where" type="adql:whereType" />

            <xs:element name="Search">
                <xs:complexType>
                    <xs:sequence>
                        <xs:element ref="tns:Where" minOccurs="1" maxOccurs="1" />
                        <xs:element name="from" type="xs:positiveInteger"
                                    minOccurs="0" maxOccurs="1" />
                        <xs:element name="max" type="xs:positiveInteger"
                                    minOccurs="0" maxOccurs="1" />
                        <xs:element name="identifiersOnly" type="xs:boolean"
                                    minOccurs="0" maxOccurs="1" />
                    </xs:sequence>
```

```
            </xs:complexType>
        </xs:element>

        <xs:element name="SearchResponse">
          <xs:complexType>
            <xs:sequence>
              <xs:element ref="ri:VOResources" minOccurs="1" maxOccurs="1" />
            </xs:sequence>
          </xs:complexType>
        </xs:element>

        <xs:element name="GetResource">
          <xs:complexType>
            <xs:sequence>
              <xs:element name="identifier" type="xs:string"
                          minOccurs="1" maxOccurs="1" />
            </xs:sequence>
          </xs:complexType>
        </xs:element>

        <xs:element name="ResolveResponse">
          <xs:complexType>
            <xs:sequence>
              <xs:element ref="ri:Resource"
                          minOccurs="1" maxOccurs="1" />
            </xs:sequence>
          </xs:complexType>
        </xs:element>

        <xs:element name="GetIdentity" />

        <xs:element name="XQuerySearch">
          <xs:complexType>
            <xs:sequence>
              <xs:element name="xquery" type="xs:string"
                          minOccurs="1" maxOccurs="1" />
            </xs:sequence>
          </xs:complexType>
        </xs:element>

        <xs:element name="XQuerySearchResponse">
          <xs:complexType>
            <xs:sequence>
              <xs:any minOccurs="0"/>
            </xs:sequence>
          </xs:complexType>
        </xs:element>

        <xs:element name="KeywordSearch">
          <xs:complexType>
            <xs:sequence>
              <xs:element name="keywords" type="xs:string"
                          minOccurs="1" maxOccurs="1" />
              <xs:element name="orValues" type="xs:boolean"
                          minOccurs="1" maxOccurs="1" />
              <xs:element name="from" type="xs:positiveInteger"
                          minOccurs="0" maxOccurs="1" />
              <xs:element name="max" type="xs:positiveInteger"
                          minOccurs="0" maxOccurs="1" />
              <xs:element name="identifiersOnly" type="xs:boolean"
                          minOccurs="0" maxOccurs="1" />
            </xs:sequence>
```

```
            </xs:complexType>
        </xs:element>

        <xs:element name="ErrorResponse">
          <xs:complexType>
            <xs:sequence>
              <xs:element name="errorMessage" type="xs:string"
                           minOccurs="1" maxOccurs="1" />
            </xs:sequence>
          </xs:complexType>
        </xs:element>

        <xs:element name="UnsupportedOperation">
          <xs:complexType>
            <xs:sequence>
              <xs:element name="errorMessage" type="xs:string"
                           minOccurs="0" maxOccurs="1" />
            </xs:sequence>
          </xs:complexType>
        </xs:element>

        <xs:element name="NotFound">
          <xs:complexType>
            <xs:sequence>
              <xs:element name="errorMessage" type="xs:string"
                           minOccurs="0" maxOccurs="1" />
            </xs:sequence>
          </xs:complexType>
        </xs:element>

    </xs:schema>
</types>

<message name="empty"/>

<message name="ErrorResp">
    <part name="ErrorResp"  element="tns:ErrorResponse"/>
</message>

<message name="SearchReq">
    <part name="Search" element="tns:Search" />
</message>

<message name="SearchResp">
    <part name="VOResources" element="tns:SearchResponse" />
</message>

<message name="GetResourceReq">
    <part name="GetResource" element="tns:GetResource" />
</message>

<message name="GetIdentityReq">
    <part name="GetIdentity" element="tns:GetIdentity" />
</message>

<message name="KeywordSearchReq">
    <part name="KeywordSearch" element="tns:KeywordSearch" />
</message>

<message name="XQuerySearchReq">
    <part name="XQuerySearch" element="tns:XQuerySearch" />
</message>
```

```
<message name="XQuerySearchResp">
   <part name="XQuerySearchResp" element="tns:XQuerySearchResponse" />
</message>

<message name="OpUnsupportedResp">
   <part name="OpUnsupportedResp"  element="tns:UnsupportedOperation"/>
</message>

<message name="NotFoundResp">
   <part name="NotFoundResp"  element="tns:NotFound"/>
</message>

<message name="ResolveResp">
   <part name="Resource" element="tns:ResolveResponse" />
</message>

<portType name="RegistrySearchPortType">
   <operation name="Search">
      <input message="tns:SearchReq" />
      <output message="tns:SearchResp" />
      <fault name="SearchError" message="tns:ErrorResp"/>
   </operation>
   <operation name="KeywordSearch">
      <input message="tns:KeywordSearchReq" />
      <output message="tns:SearchResp" />
      <fault name="KeywordSearchError" message="tns:ErrorResp"/>
   </operation>

   <operation name="GetResource">
      <input message="tns:GetResourceReq" />
      <output message="tns:ResolveResp" />
      <fault name="GetResourceError" message="tns:ErrorResp"/>
      <fault name="NotFound" message="tns:NotFoundResp"/>
   </operation>

   <operation name="GetIdentity">
      <input message="tns:GetIdentityReq" />
      <output message="tns:ResolveResp" />
      <fault name="GetIdentityError" message="tns:ErrorResp"/>
   </operation>

   <operation name="XQuerySearch">
      <input message="tns:XQuerySearchReq" />
      <output message="tns:XQuerySearchResp" />
      <fault name="XQuerySearchError" message="tns:ErrorResp"/>
      <fault name="OpUnsupported" message="tns:OpUnsupportedResp"/>
   </operation>
</portType>

<binding name="RegistrySearchSOAP" type="tns:RegistrySearchPortType">
   <soap:binding style="document"
                 transport="http://schemas.xmlsoap.org/soap/http"/>

   <operation name="Search">
      <soap:operation style="document"
         soapAction="http://www.ivoa.net/wsdl/RegistrySearch/v1.0#Search"/>
      <input>
         <soap:body use="literal"
            namespace="http://www.ivoa.net/wsdl/RegistrySearch/v1.0"/>
      </input>
      <output>
```

```
              <soap:body use="literal"
                 namespace="http://www.ivoa.net/wsdl/RegistrySearch/v1.0" />
        </output>
        <fault name="SearchError">
           <soap:fault name="SearchError" use="literal"
                 namespace="http://www.ivoa.net/wsdl/RegistrySearch/v1.0"/>
        </fault>
   </operation>

   <operation name="KeywordSearch">
        <soap:operation style="document"
              soapAction=
               "http://www.ivoa.net/wsdl/RegistrySearch/v1.0#KeywordSearch"/>
        <input>
           <soap:body use="literal"
                 namespace="http://www.ivoa.net/wsdl/RegistrySearch/v1.0"/>
        </input>
        <output>
           <soap:body use="literal"
                 namespace="http://www.ivoa.net/wsdl/RegistrySearch/v1.0"/>
        </output>
        <fault name="KeywordSearchError">
           <soap:fault name="KeywordSearchError" use="literal"
                 namespace="http://www.ivoa.net/wsdl/RegistrySearch/v1.0"/>
        </fault>
   </operation>

   <operation name="GetResource">
        <soap:operation style="document"
              soapAction=
                "http://www.ivoa.net/wsdl/RegistrySearch/v1.0#GetResource" />
        <input>
           <soap:body use="literal"
                 namespace="http://www.ivoa.net/wsdl/RegistrySearch/v1.0"/>
        </input>
        <output>
           <soap:body use="literal"
                 namespace="http://www.ivoa.net/wsdl/RegistrySearch/v1.0"/>
        </output>
        <fault name="GetResourceError">
           <soap:fault name="GetResourceError" use="literal"
                 namespace="http://www.ivoa.net/wsdl/RegistrySearch/v1.0"/>
        </fault>
        <fault name="NotFound">
           <soap:fault name="NotFound" use="literal"
                 namespace="http://www.ivoa.net/wsdl/RegistrySearch/v1.0"/>
        </fault>
   </operation>

   <operation name="GetIdentity">
        <soap:operation style="document"
              soapAction=
                "http://www.ivoa.net/wsdl/RegistrySearch/v1.0#GetIdentity" />
        <input>
           <soap:body use="literal"
                 namespace="http://www.ivoa.net/wsdl/RegistrySearch/v1.0"/>
        </input>
        <output>
           <soap:body use="literal"
                 namespace="http://www.ivoa.net/wsdl/RegistrySearch/v1.0"/>
        </output>
        <fault name="GetIdentityError">
```

```
                    <soap:fault name="GetIdentityError" use="literal"
                          namespace="http://www.ivoa.net/wsdl/RegistrySearch/v1.0"/>
            </fault>
      </operation>

      <operation name="XQuerySearch">
            <soap:operation style="document"
                  soapAction=
                      "http://www.ivoa.net/wsdl/RegistrySearch/v1.0#XQuerySearch"/>
            <input>
                  <soap:body use="literal"
                          namespace="http://www.ivoa.net/wsdl/RegistrySearch/v1.0"/>
            </input>
            <output>
                  <soap:body use="literal"
                          namespace="http://www.ivoa.net/wsdl/RegistrySearch/v1.0" />
            </output>
            <fault name="XQuerySearchError">
                  <soap:fault name="XQuerySearchError" use="literal"
                          namespace="http://www.ivoa.net/wsdl/RegistrySearch/v1.0"/>
            </fault>
            <fault name="OpUnsupported">
                  <soap:fault name="OpUnsupported" use="literal"
                          namespace="http://www.ivoa.net/wsdl/RegistrySearch/v1.0"/>
            </fault>
      </operation>

   </binding>
</definitions>
```

### RegistryInterface Schema

```
<?xml version="1.0" encoding="UTF-8"?>
<xs:schema targetNamespace="http://www.ivoa.net/xml/RegistryInterface/v1.0"
            xmlns:ri="http://www.ivoa.net/xml/RegistryInterface/v1.0"
            xmlns="http://www.w3.org/2001/XMLSchema"
            xmlns:xs="http://www.w3.org/2001/XMLSchema"
            xmlns:vr="http://www.ivoa.net/xml/VOResource/v1.0"
            elementFormDefault="qualified"
            version="1.0"

   <xs:import namespace="http://www.ivoa.net/xml/VOResource/v1.0"
            schemaLocation="http://www.ivoa.net/xml/VOResource/v1.0"/>

   <xs:element name="VOResources">
      <xs:annotation>
         <xs:documentation>
           a container for one or more resource descriptions or
           identifier references to resources.
         </xs:documentation>
         <xs:documentation>
           This is used to transmitting multiple resource descriptions
           resulting from a query.
         </xs:documentation>
      </xs:annotation>

      <xs:complexType>
         <xs:sequence>
```

```
            <xs:choice>
                <xs:element ref="ri:Resource"
                            minOccurs="0" maxOccurs="unbounded"/>
                <xs:element name="identifier" type="vr:IdentifierURI"
                            minOccurs="0" maxOccurs="unbounded"/>
            </xs:choice>
        </xs:sequence>
        <xs:attribute name="from" type="xs:positiveInteger" use="required" />
        <xs:attribute name="numberReturned" type="xs:positiveInteger"
                        use="required" />
        <xs:attribute name="more" type="xs:boolean"  use="required" />
    </xs:complexType>
</xs:element>

<xs:element name="Resource" type="vr:Resource">
    <xs:annotation>
        <xs:documentation>
          a description of a single resource
        </xs:documentation>
    </xs:annotation>
</xs:element>

</xs:schema>
```

# Appendix A.2  WSDL Document for the Harvesting Interface

The RegistryHarvest interface includes the RegistryInterface schema which is defined by the listing given in Appendix A.1.

**RegistryHarvest  Interface WSDL**
```
<?xml version="1.0" encoding="UTF-8"?>
<definitions name="OAI-PMH"
            xmlns="http://schemas.xmlsoap.org/wsdl/"
            xmlns:soap="http://schemas.xmlsoap.org/wsdl/soap/"
            xmlns:xs="http://www.w3.org/2001/XMLSchema"
            xmlns:oai="http://www.openarchives.org/OAI/2.0/"
            xmlns:tns="http://www.ivoa.net/wsdl/RegistryHarvest/v1.0"
            targetNamespace="http://www.ivoa.net/wsdl/RegistryHarvest/v1.0"

    <types>
        <xs:schema xmlns="http://www.w3.org/2001/XMLSchema"
                    xmlns:tns="http://www.ivoa.net/wsdl/RegistryHarvest/v1.0"
                    targetNamespace="http://www.ivoa.net/wsdl/RegistryHarvest/v1.0">

            <xs:import namespace="http://www.openarchives.org/OAI/2.0/"
                        schemaLocation="OAI.xsd"/>

            <xs:element name="resumptionToken" type="xs:string"/>

            <xs:element name="Identify" />
```

```
<xs:element name="IdentifyResponse">
  <xs:complexType>
    <xs:sequence>
      <xs:element minOccurs="1" maxOccurs="1" ref="oai:OAI-PMH"/>
    </xs:sequence>
  </xs:complexType>
</xs:element>

<xs:element name="ErrorResponse">
  <xs:complexType>
    <xs:sequence>
      <xs:element minOccurs="1" maxOccurs="1" ref="oai:OAI-PMH"/>
    </xs:sequence>
  </xs:complexType>
</xs:element>

<xs:element name="ListMetadataFormats">
  <xs:complexType>
    <xs:sequence>
      <xs:element name="identifier" type="xs:anyURI"
                  minOccurs="0" maxOccurs="1"/>
    </xs:sequence>
  </xs:complexType>
</xs:element>

<xs:element name="ListMetadataFormatsResponse">
  <xs:complexType>
    <xs:sequence>
      <xs:element minOccurs="1" maxOccurs="1" ref="oai:OAI-PMH"/>
    </xs:sequence>
  </xs:complexType>
</xs:element>

<xs:element name="ListSets">
  <xs:complexType>
    <xs:sequence>
      <xs:element ref="tns:resumptionToken" minOccurs="0"/>
    </xs:sequence>
  </xs:complexType>
</xs:element>

<xs:element name="ListSetsResponse">
  <xs:complexType>
    <xs:sequence>
      <xs:element minOccurs="1" maxOccurs="1" ref="oai:OAI-PMH"/>
    </xs:sequence>
  </xs:complexType>
</xs:element>

<xs:element name="GetRecord">
  <xs:complexType>
    <xs:sequence>
      <xs:element name="identifier" type="xs:anyURI" />
      <xs:element name="metadataPrefix" type="xs:string" />
    </xs:sequence>
  </xs:complexType>
</xs:element>

<xs:element name="GetRecordResponse">
  <xs:complexType>
    <xs:sequence>
```

```
                    <xs:element minOccurs="1" maxOccurs="1" ref="oai:OAI-PMH"/>
                 </xs:sequence>
              </xs:complexType>
           </xs:element>

           <xs:element name="ListIdentifiers">
              <xs:complexType>
                  <xs:sequence>
                     <xs:group ref="tns_schema:exclusiveArgs"/>
                  </xs:sequence>
              </xs:complexType>
           </xs:element>

           <xs:element name="ListIdentifiersResponse">
            <xs:complexType>
               <xs:sequence>
                  <xs:element minOccurs="1" maxOccurs="1" ref="oai:OAI-PMH"/>
               </xs:sequence>
            </xs:complexType>
           </xs:element>

           <xs:element name="ListRecords">
              <xs:complexType>
                  <xs:sequence>
                     <xs:group ref="tns_schema:exclusiveArgs"/>
                  </xs:sequence>
              </xs:complexType>
           </xs:element>

           <xs:element name="ListRecordsResponse">
             <xs:complexType>
               <xs:sequence>
                  <xs:element minOccurs="l" maxOccurs="1" ref="oai:OAI-PMH"/>
               </xs:sequence>
             </xs:complexType>
           </xs:element>

           <xs:group name="exclusiveArgs">
              <xs:choice>
                 <xs:element ref="tns_schema:resumptionToken"/>
                 <xs:sequence>
                   <xs:element name="from" type="oai:UTCdatetimeType"
                                minOccurs="0"/>
                   <xs:element name="until" type="oai:UTCdatetimeType"
                                minOccurs="0"/>
                   <xs:element name="set" type="oai:setSpecType" minOccurs="0"/>
                   <xs:element name="metadataPrefix"
                                type="oai:metadataPrefixType"/>
                 </xs:sequence>
              </xs:choice>
           </xs:group>
        </xs:schema>
</types>

<message name="ErrorResp">
    <part name="ErrorResp"  element="tns:ErrorResponse"/>
</message>

<message name="IdentifyResp">
    <part name="IdentifyResponse" element="tns:IdentifyResponse"/>
</message>
```

```xml
<message name="ListMetadataFormatsResp">
    <part name="ListMetadataFormatsResponse"
          element="tns:ListMetadataFormatsResponse"/>
</message>

<!-- ListSets verb -->
<!-- the ListSets operations take no inputs -->

<message name="ListSetsResp">
    <part name="ListSetsResponse" element="tns:ListSetsResponse"/>
</message>

<message name="ListIdentifiersResp">
    <part name="ListIdentifiersResponse"
          element="tns:ListIdentifiersResponse"/>
</message>

<!-- GetRecord verb -->
<message name="GetRecordReq">
    <part name="GetRecord" element="tns:GetRecord" />
</message>

<message name="GetRecordResp">
    <part name="GetRecordResponse" element="tns:GetRecordResponse"/>
</message>

<!-- ListIdentifiers verb -->
<message name="ListIdentifiersReq">
    <part name="ListIdentifiers" element="tns:ListIdentifiers" />
</message>

<!-- ListRecords verb -->
<message name="ListRecordsReq">
    <part name="ListRecords" element="tns:ListRecords" />
</message>

<message name="ListRecordsResp">
    <part name="ListRecordsResponse" element="tns:ListRecords"/>
</message>

<message name="ListSetsReq">
    <part name="ListSets" element="tns:ListSets" />
</message>

<message name="IdentifyReq">
    <part name="ListSets" element="tns:Identify" />
</message>

<message name="ListMetadataFormatsReq">
    <part name="ListMetadataFormats" element="tns:ListMetadataFormats" />
</message>

<portType name="RegistryHarvestPortType">

    <!-- Identify verb -->
    <operation name="Identify">
        <input message="tns:IdentifyReq"/>
        <output message="tns:IdentifyResp"/>
        <fault name="IdentifyError" message="tns:ErrorResp"/>
    </operation>

    <!-- ListMetadataFormats verb -->
```

```
      <operation name="ListMetadataFormats">
         <input message="tns:ListMetadataFormatsReq"/>
         <output message="tns:ListMetadataFormatsResp"/>
         <fault name="ListMetadataFormatsError" message="tns:ErrorResp"/>
      </operation>

      <!-- ListSets verb (with resume version) -->
      <operation name="ListSets">
         <input message="tns:ListSetsReq"/>
         <output message="tns:ListSetsResp"/>
         <fault name="ListSetsError" message="tns:ErrorResp"/>
      </operation>

      <!-- GetRecord verb -->
      <operation name="GetRecord">
         <input message="tns:GetRecordReq"/>
         <output message="tns:GetRecordResp"/>
         <fault name="GetRecordError" message="tns:ErrorResp"/>
      </operation>

      <!-- ListIdentifiers verb (with resume version) -->
      <operation name="ListIdentifiers">
         <input message="tns:ListIdentifiersReq"/>
         <output message="tns:ListIdentifiersResp"/>
         <fault name="ListIdentifiersError" message="tns:ErrorResp"/>
      </operation>

      <!-- ListRecords verb (with resume version) -->
      <operation name="ListRecords">
         <input message="tns:ListRecordsReq"/>
         <output message="tns:ListRecordsResp"/>
         <fault name="ListRecordsError" message="tns:ErrorResp"/>
      </operation>

   </portType>

   <binding name="RegistryHarvestSOAP" type="tns:RegistryHarvestPortType">
      <soap:binding style="document"
                    transport="http://schemas.xmlsoap.org/soap/http"/>

      <!-- Identify verb -->
      <operation name="Identify">
         <soap:operation
            soapAction="http://www.ivoa.net/wsdl/RegistryInterface#Identify"/>
         <input>
            <soap:body use="literal"
                       namespace="http://www.ivoa.net/wsdl/RegistryInterface"/>
         </input>
         <output>
            <soap:body use="literal"
                       namespace="http://www.ivoa.net/wsdl/RegistryInterface"/>
         </output>
         <fault name="IdentifyError">
            <soap:fault name="IdentifyError" use="literal"
                        namespace="http://www.ivoa.net/wsdl/RegistryInterface"/>
         </fault>
      </operation>

      <!-- ListMetadataFormats verb -->
      <operation name="ListMetadataFormats">
         <soap:operation
soapAction="http://www.ivoa.net/wsdl/RegistryInterface#ListMetadataFormats"/>
```

```
      <input>
         <soap:body use="literal"
                    namespace="http://www.ivoa.net/wsdl/RegistryInterface"/>
      </input>
      <output>
         <soap:body use="literal"
                    namespace="http://www.ivoa.net/wsdl/RegistryInterface"/>
      </output>
      <fault name="ListMetadataFormatsError">
         <soap:fault name="ListMetadataFormatsError" use="literal"
                    namespace="http://www.ivoa.net/wsdl/RegistryInterface"/>
      </fault>
   </operation>

   <!-- ListSets verb (with resume version) -->
   <operation name="ListSets">
      <soap:operation
         soapAction="http://www.ivoa.net/wsdl/RegistryInterface#ListSets"/>
      <input>
         <soap:body use="literal"
                    namespace="http://www.ivoa.net/wsdl/RegistryInterface"/>
      </input>
      <output>
         <soap:body use="literal"
                    namespace="http://www.ivoa.net/wsdl/RegistryInterface"/>
      </output>
      <fault name="ListSetsError">
         <soap:fault name="ListSetsError" use="literal"
                    namespace="http://www.ivoa.net/wsdl/RegistryInterface"/>
      </fault>
   </operation>

   <!-- GetRecord verb -->
   <operation name="GetRecord">
      <soap:operation
         soapAction="http://www.ivoa.net/wsdl/RegistryInterface#GetRecord"/>
      <input>
         <soap:body use="literal"
                    namespace="http://www.ivoa.net/wsdl/RegistryInterface"/>
      </input>
      <output>
         <soap:body use="literal"
                    namespace="http://www.ivoa.net/wsdl/RegistryInterface"/>
      </output>
      <fault name="GetRecordError">
         <soap:fault name="GetRecordError" use="literal"
                    namespace="http://www.ivoa.net/wsdl/RegistryInterface"/>
      </fault>
   </operation>

   <!-- ListIdentifiers verb (with resume version) -->
   <operation name="ListIdentifiers">
      <soap:operation
soapAction="http://www.ivoa.net/wsdl/RegistryInterface#ListIdentifiers"/>
      <input>
         <soap:body use="literal"
                    namespace="http://www.ivoa.net/wsdl/RegistryInterface"/>
      </input>
      <output>
         <soap:body use="literal"
                    namespace="http://www.ivoa.net/wsdl/RegistryInterface"/>
      </output>
```

```
            <fault name="ListIdentifiersError">
                <soap:fault name="ListIdentifiersError" use="literal"
                        namespace="http://www.ivoa.net/wsdl/RegistryInterface"/>
            </fault>
        </operation>

        <!-- ListRecords verb (with resume version) -->
        <operation name="ListRecords">
            <soap:operation
             soapAction="http://www.ivoa.net/wsdl/RegistryInterface#ListRecords"/>
            <input>
                <soap:body use="literal"
                        namespace="http://www.ivoa.net/wsdl/RegistryInterface"/>
            </input>
            <output>
                <soap:body use="literal"
                        namespace="http://www.ivoa.net/wsdl/RegistryInterface"/>
            </output>
            <fault name="ListRecordsError">
                <soap:fault name="ListRecordsError" use="literal"
                        namespace="http://www.ivoa.net/wsdl/RegistryInterface"/>
            </fault>
        </operation>

    </binding>
</definitions>
```

## Appendix A.3  VORegistry:  the VOResource Extension Schema for Registering Registries

**VORegistry Extension Schema**
```
<?xml version="1.0" encoding="UTF-8"?>
<xs:schema targetNamespace="http://www.ivoa.net/xml/VORegistry/v1.0"
           xmlns:xs="http://www.w3.org/2001/XMLSchema"
           xmlns:vr="http://www.ivoa.net/xml/VOResource/v1.0"
           xmlns:vg="http://www.ivoa.net/xml/VORegistry/v1.0"
           xmlns:vm="http://www.ivoa.net/xml/VOMetadata/v0.1"
           elementFormDefault="unqualified"
           attributeFormDefault="unqualified"
           version="1.0">

   <xs:import namespace="http://www.ivoa.net/xml/VOResource/v1.0"
        schemaLocation="http://www.ivoa.net/xml/VOResource/v1.0"/>

   <xs:complexType name="Registry">
      <xs:annotation>
         <xs:documentation>
            a service that provides access to descriptions of resources.
         </xs:documentation>
         <xs:documentation>
            A registry is considered a publishing registry if it
            contains a capability element with xsi:type="vg:Harvest".
            It is considered a searchable registry if it contains a
```

```
                    capability element with xsi:type="vg:Search".
            </xs:documentation>
        </xs:annotation>

        <xs:complexContent>
            <xs:extension base="vr:Service">
                <xs:sequence>
                    <xs:element name="full" type="xs:boolean">
                        <xs:annotation>
                            <xs:documentation>
                              If true, this registry attempts to collect all
                              resource records known to the IVOA.
                            </xs:documentation>
                            <xs:documentation>
                              A registry typically collects everything by
                              harvesting from all registries listed in the
                              IVOA Registry of Registries.
                            </xs:documentation>
                        </xs:annotation>
                    </xs:element>

                    <xs:element name="managedAuthority"
                                  type="vr:AuthorityID"
                                  minOccurs="0" maxOccurs="unbounded">
                        <xs:annotation>
                            <xs:documentation>
                              an authority identifier managed by the registry.
                            </xs:documentation>
                            <xs:documentation>
                              Typically, this means the AuthorityIDs that
                              originated (i.e. were first published by) this
                              registry.  Currently, only one registry can lay
                              claim to an AuthorityID via this element at a
                              time.
                            </xs:documentation>
                        </xs:annotation>
                    </xs:element>
                </xs:sequence>
            </xs:extension>
        </xs:complexContent>
</xs:complexType>

<xs:complexType name="RegCapRestriction" abstract="true">
    <xs:annotation>
        <xs:documentation>
          an abstract capability that fixes the standardID to the
          IVOA ID for the Registry standard.
        </xs:documentation>
        <xs:documentation>
          See vr:Capability for documentation on inherited children.
        </xs:documentation>
    </xs:annotation>
    <xs:complexContent>
        <xs:restriction base="vr:Capability">
            <xs:sequence>
                <xs:element name="validationLevel" type="vr:Validation"
```

```
                          minOccurs="0" maxOccurs="unbounded"/>
            <xs:element name="description" type="xs:token"
                          minOccurs="0"/>
            <xs:element name="interface" type="vr:Interface"
                          minOccurs="0" maxOccurs="unbounded"/>
        </xs:sequence>
        <xs:attribute name="standardID" type="vr:IdentifierURI"
                          use="required"
                          fixed="ivo://ivoa.net/std/Registry"/>
      </xs:restriction>
    </xs:complexContent>
</xs:complexType>

<xs:complexType name="Harvest">
    <xs:annotation>
      <xs:documentation>
        The capabilities of the Registry Harvest implementation.
      </xs:documentation>
    </xs:annotation>

    <xs:complexContent>
      <xs:extension base="vg:RegCapRestriction">
        <xs:sequence>

            <xs:element name="maxRecords" type="xs:int">
              <xs:annotation>
                <xs:documentation>
                    The largest number of records that the registry
                    search method will return.  A value greater
                    than one implies that an OAI continuation token
                    will be provided when the limit is reached.  A
                    value of zero or less indicates that there is
                    no explicit limit and thus, continuation tokens
                    are not supported.
                </xs:documentation>
              </xs:annotation>
            </xs:element>

        </xs:sequence>
      </xs:extension>
    </xs:complexContent>
</xs:complexType>

<xs:complexType name="Search">
    <xs:annotation>
      <xs:documentation>
        The capabilities of the Registry Search implementation.
      </xs:documentation>
    </xs:annotation>

    <xs:complexContent>
      <xs:extension base="vg:RegCapRestriction">
        <xs:sequence>

            <xs:element name="maxRecords" type="xs:int">
              <xs:annotation>
```

```
                    <xs:documentation>
                        The largest number of records that the registry
                        search method will return.  A value of zero or
                        less indicates that there is no explicit limit.
                    </xs:documentation>
                </xs:annotation>
            </xs:element>

            <xs:element name="extensionSearchSupport"
                        type="vg:ExtensionSearchSupport">
                <xs:annotation>
                    <xs:documentation>
                      the level of support provided for searching
                      against metadata defined in a legal VOResource
                      extension schema.
                    </xs:documentation>
                    <xs:documentation>
                      A legal VOResource extension schema is one that
                      imports and extends the VOResource core schema
                      in compliance with the VOResource standard.
                    </xs:documentation>
                </xs:annotation>
            </xs:element>

            <xs:element name="optionalProtocol"
                        type="vg:OptionalProtocol"
                        minOccurs="0" maxOccurs="unbounded">
                <xs:annotation>
                    <xs:documentation>
                      the name of an optional advanced search
                      protocol supported.
                    </xs:documentation>
                    <xs:documentation>
                        Only one optional protocol is currently allowed
                        (XQuery).  It is assumed that the required
                        protocols (simple keyword search and ADQL) are
                        supported.
                    </xs:documentation>
                </xs:annotation>
            </xs:element>

        </xs:sequence>
      </xs:extension>
    </xs:complexContent>
</xs:complexType>

<xs:simpleType name="ExtensionSearchSupport">
    <xs:restriction base="xs:NMTOKEN">
      <xs:enumeration value="core">
        <xs:annotation>
          <xs:documentation>
            Only searches against the core VOResource metadata are
            supported.
          </xs:documentation>
        </xs:annotation>
      </xs:enumeration>
```

```
            <xs:enumeration value="partial">
               <xs:annotation>
                  <xs:documentation>
                     Searches against some VOResource extension metadata
                     are supported but not necessarily all that exist in
                     the registry.
                  </xs:documentation>
               </xs:annotation>
            </xs:enumeration>
            <xs:enumeration value="full">
               <xs:annotation>
                  <xs:documentation>
                     Searches against all VOResource extension metadata
                     contained in the registry are supported.
                  </xs:documentation>
               </xs:annotation>
            </xs:enumeration>
         </xs:restriction>
</xs:simpleType>

<xs:simpleType name="OptionalProtocol">
      <xs:restriction base="xs:NMTOKEN">
         <xs:enumeration value="XQuery">
            <xs:annotation>
               <xs:documentation>
                  the XQuery (http://www.w3.org/TR/xquery/) protocol as
                  defined in the VO Registry Interface standard.
               </xs:documentation>
            </xs:annotation>
         </xs:enumeration>
      </xs:restriction>
</xs:simpleType>

<xs:complexType name="OAIHTTP">
      <xs:annotation>
         <xs:documentation>
            a description of the standard OAI PMH interface using HTTP
            (GET or POST) queries.
         </xs:documentation>
         <xs:documentation>
            the accessURL child element is the base URL for the OAI
            service as defined in section 3.1.1 of the OAI PMH
            standard.
         </xs:documentation>
      </xs:annotation>

      <xs:complexContent>
         <xs:extension base="vr:Interface">
             <xs:sequence/>
         </xs:extension>
      </xs:complexContent>
</xs:complexType>

<xs:complexType name="OAISOAP">
      <xs:annotation>
```

```
            <xs:documentation>
              a description of the standard OAI PMH interface using a SOAP
              Web Service interface.
            </xs:documentation>
            <xs:documentation>
              the accessURL child element is the service port location URL
              for the OAI SOAP Web Service.
            </xs:documentation>
        </xs:annotation>

        <xs:complexContent>
            <xs:extension base="vr:WebService">
                <xs:sequence/>
            </xs:extension>
        </xs:complexContent>

    </xs:complexType>

    <xs:complexType name="Authority">
        <xs:annotation>
            <xs:documentation>
              a naming authority; an assertion of control over a
              namespace represented by an authority identifier.
            </xs:documentation>
        </xs:annotation>
        <xs:complexContent>
            <xs:extension base="vr:Resource">
                <xs:sequence>

                    <xs:element name="managingOrg" type="vr:ResourceName">
                        <xs:annotation>
                            <xs:documentation>
                              the organization that manages or owns this
                              authority.
                            </xs:documentation>
                            <xs:documentation>
                              In most cases, this will be the same as the
                              Publisher.
                            </xs:documentation>
                        </xs:annotation>
                    </xs:element>

                </xs:sequence>
            </xs:extension>
        </xs:complexContent>
    </xs:complexType>

</xs:schema>
```

## Appendix A.4 IVOA Recommended Extension Prefixes

Section 3.1.2 strongly recommends the consistent use of namespace prefixes across all compliant registries.  The table below provides the recommend list of prefixes for the schemas that are commonly used in VOResource records as of this writing.  For IVOA standard schemas not listed here, the prefix recommended by the standard defining the schema should be used.

| Prefix | Namespace |
|--------|-----------|
| sia | http://www.ivoa.net/xml/SIA/v### |
| cs | http://www.ivoa.net/xml/ConeSearch/v### |
| vs | http://www.ivoa.net/xml/VODataService/v### |
| ssa | http://www.ivoa.net/xml/SSA/v### |
| vg | http://www.ivoa.net/xml/VORegistry/v### |
| vstd | http://www.ivoa.net/xml/VOStandard/v### |
| va | http://www.ivoa.net/xml/VOApplication/v### |
| cea | http://www.ivoa.net/xml/CEA/v### |
| stc | http://www.ivoa.net/xml/STC/### |
| sn | http://www.ivoa.net/xml/SkyNode/v### |

**Note:**
Schemas found at the http://www.ivoa.net/xml/ will have annotations indicating the recommended prefix.  The recommended prefix can also be found in VOResource record describing the standard in the Registry of Registries [RofR].

# Appendix A.5 ADQL for Querying Registries

When the registry search interface was first developed for standardization, it was intended that it would build on an existing standard for ADQL.  In particular, the search interface was based on ADQL v1.01 [ADQL].  This ADQL specification never evolved out of the Working Draft status.  It was eventually superseded by v2.0 which was sufficiently different from v1.01 that RI could not be revised easily.  Instead, RI continues to be based on ADQL v1.01.

In lieu of a reference to an actual IVOA Recommendation for ADQL, this appendix excerpts the relevant sections of the ADQL v1.01 working draft [ADQL].

## A.5.1 Introduction

The Astronomical Data Query Language (ADQL) is the language used by the International Virtual Observatory Alliance (IVOA) to represent astronomy queries posted to VO data services (SkyNodes). IVOA has developed several standardized protocols to access astronomical databases, e.g., SIAP for image data and SSAP for spectral data.  Current ADQL (ADQL 1.0) is designed to access astronomical catalog data only through the SkyNode Interfaces. The work to integrate SIAP, SSAP and ADQL is under progress toward the future version of ADQL.

ADQL 1.0 is based on Structured Query Language (SQL), especially on SQL92. The VO has a number of tabular data sets and many of them are stored in relational databases (RDBs), making SQL a convenient access means.

## A.5.2 Astronomical Data Query Language (ADQL)

ADQL is based on a subset of SQL which has been extended to support queries which are specific to astronomy.

ADQL has two forms or representations:

- **ADQL/s** : A String form based on SQL92 and conforming to the ADQL grammar in Appendix A.5.6. Some non standard SQL extensions have been added to support distributed astronomical queries; and
- **ADQL/x** : An XML document conforming to the ADQL schema (XSD) included in Section 4. The XML document is the mechanism used to pass a query to the SkyNode Web service Interface.

ADQL/s and ADQL/x are translatable to each other without loss of information.

Since ADQL is similar in semantics to SQL, the requirements below list differences or special considerations only.

### A.5.2.1 Restrictions on SQL92

The formal notation for syntax of computing languages is often expressed in the "Backus Naur Form" BNF. BNF is used by popular tools such as LEX and YACC2 for producing parsers for a given syntax. Appendix A.5.6 provides the YACC type grammar for ADQL/s.

The BNF exactly defines the form of SQL92 which is ADQL/s. In essence this is any valid SQL statement. However ADQL has restrictions described below:

- In ADQL built-in functions which are defined on the server system may be called.
- INTO is supported for future interoperability with VOSpace.
- ADQL/s comments will only be supported using the /* */ syntax to delimit comments.

**Note:**
None of these restrictions are directly relevant to the use of ADQL in RI.

### A.5.2.2 Extensions to SQL92

This specification adds requirements on top of SQL92. ADQL SHALL support the extension described below.

- All table names in ADQL MUST have an alias.
- ADQL adds a keyword REGION to be used in the WHERE clause to specify search constraints.
- JDBC mathematical functions shall be allowed in ADQL.
- The XMATCH function implies cross-match between two or more astronomical catalogues.
- To support XQuery as well as SQL, and since some of our data formats are described as XSD, it will be possible to express selections and selection criteria as a simple XPath. Square brackets ([,]) and standard operators such as parent are NOT supported.
- ADQL supports the TOP syntax to return only the first N records from a query.
- ADQL allows units for all constant values specified in the query. These are optional.
- ADQL supports the use of `[ ]` to enclose literal names which may otherwise cause parse error.
- ADQL/s shall support the region keyword.

> **Note:**
> Except for the one regarding XPath identifiers, all the above-mentioned extensions are made irrelevant to RI by the restrictions enumerated in section 2.1.2.1.

### A.5.2.3 Version Information

ADQL/x documents SHALL contain a version identifier for the version of ADQL. This will start as 1.0. The version number is a dot separated string of numbers. The version number is included in the document solely so the receiving node may decide if it wishes to deal with the document or thrown an exception.

## A.5.3 ADQL Example

An ADQL/s might be as follows:

> **ADQL/s example**
> ```
> SELECT a.objid, a.ra, a.dec
> FROM SDSSDR2:Photoprimary a
> WHERE Region('CIRCLE J2000 181.3 -0.76 6.5')
> ```

This would be represented in ADQL/x as follows:

> **ADQL/x example**

```
<?xml version="1.0" encoding="utf-8"?>
<Select xmlns:xsd="http://www.w3.org/2001/XMLSchema"
        xmlns:xsi="http://www.w3.org/2001/XMLSchema-instance"
        xmlns="http://www.ivoa.net/xml/ADQL/v1.0">
  <SelectionList>
    <Item xsi:type="columnReferenceType" Table="a" Name="objid" />
    <Item xsi:type="columnReferenceType" Table="a" Name="ra" />
  </SelectionList>
  <From>
    <Table xsi:type="archiveTableType" Archive="SDSSDR2"
          Name="Photoprimary" Alias="a" />
  </From>
  <Where>
    <Condition xsi:type="regionSearchType">
      <Region xmlns:stc="http://www.ivoa.net/xml/STC/STCregion/v1.10"
            xsi:type="stc:circleType" unit="deg">
        <stc:Center>181.3 -0.76</stc:Center>
        <stc:Radius>6.5</stc:Radius>
      </Region>
    </Condition>
  </Where>
</Select>
```

## A.5.4 ADQL Grammar

This section provides an abridged version of the ADQL grammar in BNF form.
Parts of the grammar that are irrelevant to RI have been left out for brevity.

**ADQL/s Grammar**
```
sql : selectStatement (SEMICOLON)? EOF;
selectStatement : queryExpression (computeClause)? (forClause)?
(optionClause)?;
queryExpression : subQueryExpression
      (unionOperator subQueryExpression)* orderByClause)?;
subQueryExpression : querySpecification
                   | LPAREN queryExpression RPAREN;
querySpecification :
   selectClause (fromClause)? (whereClause)? (groupByClause)?
   (havingClause)? ;
selectClause :SELECT (all_distinct)? (restrictClause)? selectList;
whereClause : WHERE searchCondition;
searchCondition :
   subSearchCondition ((AND | OR) subSearchCondition)*;
subSearchCondition :
  (NOT)? (
    (LPAREN searchCondition RPAREN) => LPAREN searchCondition RPAREN
    | predicate | xMatch | region
  );
predicate :
  (
```

```
    expression
    (
     // expression comparisonOperator expression
     comparisonOperator ( expression )
     | IS (NOT)? NULL
     | (NOT)? (
         LIKE expression
         | BETWEEN expression AND expression
     )
    )
  );
region: REGION LPAREN regionClause RPAREN;
regionClause: QuotedIdentifier;
xMatch: XMATCH LPAREN xAlias (COMMA xAlias)* (
    (COMMA xSigma RPAREN ) |
    (RPAREN LESSTHAN xSigma)
    );
xSigma: number;
selectList : selectItem ( COMMA selectItem )*;
selectItem : (STAR // "*, *" is a valid select list
| (
    // starts with: "alias = column_name"
    (alias2) => (
     (alias2 dbObject COMMA) => alias2 column
     | (alias2 dbObject (binaryOperator | LPAREN)) => alias2 expression
            | (alias2 column) => alias2 column
            | (alias2 expression) => alias2 expression
            )
    // all table columns: "table.*"
    | (tableColumns) => tableColumns
    | (explicitFunction) => explicitFunction
    | (function) => function
    // some shortcuts:
    | (dbObject (alias1)? COMMA) => column (alias1)?
    | (dbObject (binaryOperator | LPAREN) ) => expression (alias1)?
    | (column) => column (alias1)?
    | (expression) => expression (alias1)?
  )
);
fromClause : FROM tableSource (COMMA tableSource)*;
tableSource : subTableSource (joinedTable)*;
subTableSource :
  (
   LPAREN (
       (joinedTables) => joinedTables RPAREN
       | (queryExpression) => queryExpression RPAREN alias1
     )
   | (function) => function (alias1)?
   | (archiveTable)? dbObject (alias1)?
       ( (WITH)? LPAREN tableHint (COMMA tableHint)* RPAREN )?
  | Variable (alias1)?
  | COLON COLON function (alias1)? // built-in function
);
dbObject : (identifier | IDENTITYCOL | ROWGUIDCOL
              | keywordAsIdentifier)
    (
```

```
      DOT (identifier | IDENTITYCOL | ROWGUIDCOL | keywordAsIdentifier)
     | (DOT DOT) => DOT DOT (identifier | IDENTITYCOL | ROWGUIDCOL
| keywordAsIdentifier)
     )*;
stringLiteral : UnicodeStringLiteral | ASCIIStringLiteral;
identifier: NonQuotedIdentifier | QuotedIdentifier;
constant : Integer | Real | NULL | stringLiteral | HexLiteral
         | Currency | ODBCDateTime | systemVariable;
unaryOperator : PLUS | MINUS | TILDE;
binaryOperator : arithmeticOperator | bitwiseOperator;
arithmeticOperator : (PLUS | MINUS | STAR | DIVIDE | MOD);
bitwiseOperator : AMPERSAND | TILDE | BITWISEOR | BITWISEXOR;
comparisonOperator :
 (
     ASSIGNEQUAL | NOTEQUAL1 | NOTEQUAL2 | LESSTHANOREQUALTO1
   | LESSTHANOREQUALTO2 | LESSTHAN | GREATERTHANOREQUALTO1
   | GREATERTHANOREQUALTO2 |GREATERTHAN
);
logicalOperator :  ALL | AND | ANY | BETWEEN | EXISTS | IN | LIKE
               | NOT | OR | SOME;
unionOperator : UNION (ALL)?;
number: (SIGN)? Number;
DOT: '.';
COLON : ':' ;
COMMA : ',' ;
SEMICOLON : ';' ;
LPAREN : '(' ;
RPAREN : ')' ;
ASSIGNEQUAL : '=' ;
NOTEQUAL1 : "<>" ;
NOTEQUAL2 : "!=" ;
LESSTHANOREQUALTO1 : "<=" ;
LESSTHANOREQUALTO2 : "!>" ;
LESSTHAN : "<" ;
GREATERTHANOREQUALTO1 : ">=" ;
GREATERTHANOREQUALTO2 : "!<" ;
GREATERTHAN : ">" ;
DIVIDE : '/' ;
PLUS : '+' ;
MINUS : '-' ;
STAR : '*' ;
MOD : '%' ;
AMPERSAND : '&' ;
TILDE : '~' ;
BITWISEOR : '|' ;
BITWISEXOR : '^' ;
DOT_STAR : ".*" ;
NOT : '!';
QUESTIONMARK : '?';
Whitespace : (' ' | '\t' | '\n' | '\r');
Letter : 'a'..'z' | '_' | '#' | '@' | '\u0080'..'\ufffe';
Digit : '0'..'9';
Integer :;
Real :;
Exponent : 'e' (Sign)? (Digit)+ ;
Sign : (PLUS | MINUS);
```

```
Number :
    ( (Digit)+ ('.' | 'e') ) =>
                (Digit)+ ( '.' (Digit)* (Exponent)? | Exponent)
    | '.' { _ttype = DOT; } ( (Digit)+ (Exponent)? )?
    | (Digit)+ { _ttype = Integer; }
    | "0x" ('a'..'f' | Digit)* ;
```

## A.5.5 ADQL XSD

The XML schema for ADQL/x is found at http://www.ivoa.net/xml/ADQL/ADQL-v1.0.xsd.  The data model for this schema is based on the ADQL Grammar presented in Appendix A.5.4.